\documentclass[usenatbib]{mn2e}
\usepackage[dvips]{graphicx}
\title[Are First Brightest Galaxies Special?]{The Non-Parametric Model for Linking 
Galaxy Luminosity with Halo/Subhalo Mass: Are First Brightest Galaxies Special?}
\author[A. Vale and J. P. Ostriker]{A. Vale$^{1,2}$\thanks{E-mail: avale@fisica.ist.utl.pt} and J. P. Ostriker$^{1,3}$\\
$^{1}$Institute of Astronomy, University of Cambridge, Madingley Road,
Cambridge CB3 0HA, United Kingdom\\
$^{2}$CENTRA, Departamento de F\'\i sica, Instituto Superior T\'ecnico, Av. Rovisco Pais 1, 1049-001 Lisboa, Portugal\\
$^{3}$Princeton University Observatory, Princeton University,
Princeton NJ 08544, USA}
\date{\today}

\begin{document}
\maketitle

\begin{abstract}

We revisit the longstanding question of whether first brightest
cluster galaxies are statistically drawn from the same distribution as
other cluster galaxies or are ``special'',
using the new non-parametric, empirically based, model presented in
\citet{paper2} for associating galaxy luminosity with halo/subhalo
masses.

We introduce scatter in galaxy luminosity at fixed halo mass into this
model, building a conditional luminosity function (CLF) by considering
two possible models: a simple lognormal and a model based on the distribution
of concentration in haloes of a given mass. 
We show that this model naturally allows an identification of
halo/subhalo systems with groups and clusters of galaxies, giving rise
to a clear central/satellite galaxy distinction, obtaining a special
distribution for the brightest cluster galaxies (BCGs).

Finally, we use these results to build up the dependence of BCG
magnitudes on cluster luminosity, focusing on two statistical
indicators, the dispersion in BCG magnitude and the magnitude
difference between first and second brightest galaxies. We compare our
results with two simple models for BCGs: a statistical hypothesis that
the BCGs are drawn from a universal distribution, and a cannibalism
scenario merging two galaxies from this distribution. The statistical
model is known to fail from work as far back as \citet{tr}. We show
that neither the statistical model nor the simplest possibility of
cannibalism provide a good match for observations, while a more 
realistic cannibalism scenario works better. Our CLF models both give
similar results, in good agreement with observations. 
Specifically, we find $<m_1>$ between -25 and -25.5 in the K-band,
$\sigma(m_1)\sim0.25$ and $<\Delta_{12}>$ between 0.6 and 0.8, for
cluster luminosities in the range of $10^{12}$ to $10^{13} h^{-2} 
{\rm L_\odot}$.

\end{abstract}


\section{Introduction}

The nature of brightest cluster galaxies (BCGs) has long been a
subject of interest and much debate
\citep{peebles,sandage,dressler}. In particular, investigators have
asked whether their origin is statistical or special in nature, that
is, whether they follow a special distribution independent of the
fainter galaxies in the cluster, or on the contrary, they are merely
the extreme values of the same global distribution derived for all
cluster galaxies.

On the theoretical side, there has been renewed interest in this 
subject with recent 
studies of the relation between galaxies and their dark matter haloes
from a theoretical, statistical point of view, involving the study of
the distribution of the galaxy population through different haloes
while bypassing the complications of the physics of galaxy
formation (e.g., \citealt{bg,paper1,iro,yanghod,zehavi,zz,coorayc,charlie,vdb}). 
Since these involve populating dark matter haloes with
galaxies, they usually lead to a distinction between central and
satellite galaxies. This
in turn has lead to, in many of these works, central galaxies being
treated separately from the rest, and therefore having a distinct
distribution, with consequences visible, for example, in the
luminosity function. Some of these studies have in fact looked
at some specific BCG-related properties of clusters, like 
the magnitude gap (e.g., \citealt{milos,vdb}).

In the past, observational studies which have focused on this issue
\citep{tr,hgt,sgh,bb84,hs,bhavsar,pl,bb} have been hindered by the limited
numbers of high luminosity galaxy observations available, since the
strongly declining nature of the bright end of the luminosity function
requires having very large samples to obtain
significant numbers of high luminosity galaxies. Due to this, these
studies were mostly inconclusive when it came to answer the question
of whether BCGs were statistical or special in nature, although many
works hinted at the latter. More recently, the advent of large scale
surveys such as the 2dF Galaxy Redshift Survey or the Sloan Digital Sky
Survey (SDSS), has motivated plentiful, ongoing work on this subject
(e.g., \citealt{lms,lm,ls,bernardi,linden}).

This issue is in large part motivated by the fact that,
observationally, BCGs do look different from other galaxies. They 
usually sit at the centre of the cluster, and tend to be
considerably brighter than the remaining cluster members. The most
striking case is cD galaxies, found in the centre of rich clusters and
which dominate their satellites in both size and brightness, while
having a characteristically distinct morphology and surface brightness
distribution (e.g., \citealt{cdreview}). Likewise, cD galaxies tend to
be brighter than what would be expected from the bright end of the
cluster galaxy luminosity function. In fact, it has been observed
that, when analyzing composite luminosity functions of cluster
galaxies, the most luminous of them form a hump at the bright end
(e.g., \citealt{colless,yagi,2pigg}). Yet, at the same time, there is
little variation in magnitude among them \citep{hs,pl,bernardi,linden}. This ties in
with the fact that the luminosity of BCGs is expected to vary only
slowly with increasing cluster luminosity (\citealt{lm}; see also the
results for the mass luminosity relation of central galaxies in
\citealt{paper2}).

In order to try to answer this problem from available observational
data, two different indicators have been considered. One is
the shape of the overall distribution of the magnitude of BCGs. If
BCGs are merely the extreme cases of a general distribution applicable
to all cluster galaxies, then it is expected that results from extreme value
theory in statistics apply, predicting a resulting distribution shaped
like the Gumbel distribution \citep{bb84,bb}. On the other hand, if
BCGs are considered a special, distinct type of galaxy, then some
particular distribution is to be expected, such as a Gaussian
\citep{pl} or lognormal. Some studies have also raised the possibility that it
could be actually a combination of the two, probably depending on the
type of cluster \citep{bhavsar,bb}.

The other property studied is the ratio $r=\Delta_{12}/\sigma_1$,
where $\Delta_{12}$ is the average magnitude difference between the
first and second brightest galaxies, and $\sigma_1$ the dispersion in
the magnitude of the first brightest galaxy. It is possible to prove
the powerful conclusion \citep{tr} that, if all galaxies are drawn
from the same statistical distribution, regardless of its exact form,
then $r\leq 1$.  Observational results give a value for $r$ around 1.5
(e.g., \citealt{lm,ls}), which would exclude this possibility.

This has led to the study of possible alternative scenarios for the
formation of BCGs, in order to account for their special nature. One
such is galactic cannibalism, initially proposed by Ostriker and
collaborators \citep{ot,oh,ho}. Such a scenario is akin to taking the
above case of having all galaxies drawn from the same distribution,
but then merging the brightest of them with one or more of the
others. From this simplistic model of the process, it is easy to see that this
mechanism would help to solve the above problem, mostly by increasing
the value of $\Delta_{12}$ as the luminosity of the first brightest
galaxy is driven up by the mergers and the brightness of the surviving
second brightest galaxy declines as luminous galaxies are merged
out of existence.

In the present paper, we explore this issue in light of the
non-parametric model for the mass luminosity relation presented in
\citet{paper2} (hereafter paper I; see also \citealt{paper1}). The
basic idea behind the non-parametric model is to adopt the simple
proposition that more luminous galaxies are hosted in more massive
haloes/subhaloes. No attempt at physical modelling is made and the
association is made simply by matching one-to-one the rank ordered
observational list of galaxies with the rank ordered computed list of
haloes/subhaloes. We here extend this model by introducing scatter
into it, and also by considering possible effects on the total
disruption of some subhaloes into the total luminosity related to the
halo.

As is the case in HOD
models (e.g., \citealt{bg,zehavi,zz}), this model naturally gives rise
to a separation between central and satellite galaxies, by associating
the former with the parent halo itself and the latter with the
subhaloes associated with it. We analyze this issue in more detail,
studying how it affects the cluster galaxy luminosity function and
gives rise to a bright end bump caused by the central galaxies. We
then develop a model for the BCG luminosity distribution. Since the
halo in fact arises from the union of subhaloes this non-parametric
model is a statistically well defined variant of the cannibalism
scenario.

This paper is organised as follows: in section 2, we give a brief
summary of the non-parametric model relating galaxy luminosity with
halo/subhalo mass presented in paper I, and introduce a simple recipe
for checking the contribution of destroyed subhaloes to the halo 
mass, and how this changes our estimate of the total luminosity. 
In section 3, we introduce scatter into the non-parametric model
by building a conditional luminosity function, where we consider
two possibilities for it, either a simple lognormal shape or 
a better motivated approach involving the distribution of concentration
for haloes of a given mass. In section 4, we explore more
indepth how the model gives rise to a central/satellite galaxy
separation, and show how this impacts the cluster galaxy luminosity
function. In section 5, we build up a model for the distribution of
cluster galaxies, based on the mass-luminosity
relation and the halo/subhaloes separation which underpins it. In
section 6, we present simple models to account for another two
possible origins for the BCG distribution: first, we consider that all
cluster galaxies are drawn from the same distribution; then we take a
simple model for cannibalism, by merging two of the galaxies (the
brightest plus one other) in the first example. Finally, in section 7
we present the results of all models for the average magnitude
of first and second brightest galaxies as well the dispersion of the
former as a function of cluster luminosity. We then
compare these results with observations.

Throughout we have used a concordance cosmological model, with
$\Omega_m=0.24$, $\Omega_\Lambda=0.76$, $h=0.735$ and $\sigma_8=0.74$ 
\citep{wmap}.

\section{The Mass-luminosity relation}

The work presented below is based on the non-parametric model for
relating galaxy luminosity with halo/subhalo mass presented in
paper I. The basic idea is that more massive haloes/subhaloes
have deeper potential wells and will thus accrete more gas and
subsequently will have more luminous galaxies forming within them. In
effect, we take the relation between galaxy luminosity and
halo/subhalo mass to be one to one and monotonic. An additional extra
ingredient is necessary to maintain this approximation in the
framework of the model, since subhaloes lose mass to the parent halo
after accretion due to tidal interactions. Alternatively put, a halo
is not simply the sum of the identifiable subhaloes within it due to 
tidal stripping. Therefore, we need to
account for the mass of the subhaloes not at present, but that which
they had at the time of their merger into the parent. The relation
between mass and luminosity is then obtained statistically by matching
the numbers of galaxies with the total number of hosts, that is,
haloes plus subhaloes through their distributions.

The halo abundance is given by the usual Sheth-Tormen mass function
\citep{stmf}:

\begin{equation} \label{stmf}
n_h(M) dM = A \Big( 1+\frac{1}{\nu^{2q}}\Big) \sqrt{\frac{2}{\pi}} 
\frac{\rho_m}{M} \frac{d\nu}{dM} {\rm exp}\Big(-\frac{\nu^2}{2}\Big) dM\, ,
\end{equation}

\noindent with $\nu=\sqrt{a}\frac{\delta_c}{D(z) \sigma(M)}$,
$a=0.707$, $A\approx 0.322$ and $q=0.3$; as usual, $\sigma(M)$ is the
variance on the mass scale $M$, $D(z)$ is the growth factor, and
$\delta_c$ is the linear threshold for spherical collapse, which in
the case of a flat universe is $\delta_c=1.686$, with a small correction
dependent on $\Omega_m$ ($\delta_c=1.673$ for $\Omega_m=0.24$).

Following the discussion in paper I, we will assume a very
simple model for the subhalo mass distribution within the parent. In
terms of their original, pre-accretion mass, we assume that the
subhalo distribution is given by a simple Schechter function:

\begin{equation} \label{shmf}
N(m|M) dm = A(M) (m/\beta M)^{-\alpha} {\rm exp} (-m/\beta M) dm/\beta M \, ,
\end{equation}

\noindent where the cutoff parameter $\beta=0.5$ serves to insure that
no subhalo was larger than half the present mass of the parent
(otherwise it would, by definition, be the parent). The slope
$\alpha=1.9$ is set to the same value as is generally found for the
present day subhalo mass function in simulations 
(e.g., \citealt{gao, jochen,vdbshmf,zentner,laurie}). 
The normalization $A(M)=1/\beta[\Gamma(2-\alpha)-\Gamma(2-\alpha,1)]$ is
set so that the total mass originally in subhaloes corresponds to the
present day mass (where the integration is done to an upper limit of $0.5M$). 
This approximation potentially ignores the problem of total disruption of
some of the merged subhaloes, as can occur for example in the case of 
major mergers, by assuming that all of these subhaloes are still present
and that therefore the total fraction of mass originally in subhaloes is one.
In \citet{paper2}, we showed that as long as this fraction is close to one, 
then the resulting mass luminosity relation is similar, with both number of
satellites in a halo and their total luminosity decreasing slightly. From the 
study of simulation results it is still not completely clear how to treat 
this complex issue, and no simple
analytical models are available, so we
explore a simple recipe to better account for this problem in the context of
our model in section \ref{sect:destroyedsh}. 

The galaxy distribution is given by the luminosity function. This is
given by the usual Schechter function fit:

\begin{equation} \label{schechter}
\phi_{obs}(L) dL = \phi_* \Big(\frac{L}{L_*}\Big)^{\alpha} {\rm
exp}\Big(-\frac{L}{L_*}\Big) \frac{dL}{L_*} \, .
\end{equation}

\noindent The values of the parameters will depend on the waveband
used. In this paper, we use mostly the K-band luminosity function from
the 2MASS survey, with parameters given by $\alpha=-1.09$,
$\phi_*=1.16\times10^{-2} h^{3} {\rm Mpc^{-3}}$ and $M_*-5 {\rm log}
h=-23.39$ \citep{2mass}. For comparison, we also obtain the
$b_J$-band 2dF survey, with $\alpha=-1.21$, $\phi_*=1.61\times10^{-2}
h^{3} {\rm Mpc^{-3}}$ and $M_*-5 {\rm log} h=-19.66$ \citep{2df}. Also
note that we are in fact extending these fits as necessary, including
beyond the magnitude interval in which they were obtained.

The basic mass-luminosity relation can then be obtained from these
ingredients by a counting process, matching the numbers of galaxies at a given luminosity
to the total number of hosts at a given mass:

\begin{equation} \label{mlrel}
\int_L^\infty \phi(L)dL=\int_M^\infty(n_h(M)+n_{sh}(M))dm \, ,
\end{equation}

\noindent where the host contribution is separated into a halo term,
$n_h(M)$, and a subhalo term obtained by summing up all the subhaloes
at that mass, $n_{sh}(m)=\int_0^\infty N(m|M)n_h(M)dM$. An average
relation between host mass and galaxy luminosity can then be built
through this process, with results that match well with observations
(see paper I for a detailed analysis). The resulting relation can be well 
fit by a double power law of the type:

\begin{equation} \label{mlfit}
L_{ref}(M)=L_0\frac{(M/M_0)^a}{[1+(M/M_0)^{b k}]^{1/k}} \, ,
\end{equation}

\noindent where the differents parameters are shown in table
\ref{mlfitparam}; mass is in units of $h^{-1} {\rm M}_\odot$,
luminosity in $h^{-2} {\rm L}_\odot$. The fit was done in the mass
range $10^{11}$ ($3\times 10^{10}$ in the $b_j$ band case) to
$3\times10^{15}$ $h^{-1} {\rm M}_\odot$.

\begin{table} \label{mlfitparam}
\begin{center}
\begin{tabular}[c]{|l|c|c|}
  & K-band & $b_J$-band \\
\hline
$L_0$ & $1.37\times10^{10}$ & $4.12\times 10^9$\\
$M_0$ & $6.14\times10^9$ & $1.66\times 10^{10}$\\
$a$ & 21.03 & 6.653 \\
$b$ & 20.74 & 6.373 \\
$k$ & 0.0363 & 0.111 \\
\end{tabular}
\caption{Fit parameters for the mass-luminosity relation.}
\end{center}
\end{table}

\begin{figure}
\includegraphics[width=84mm,angle=270]{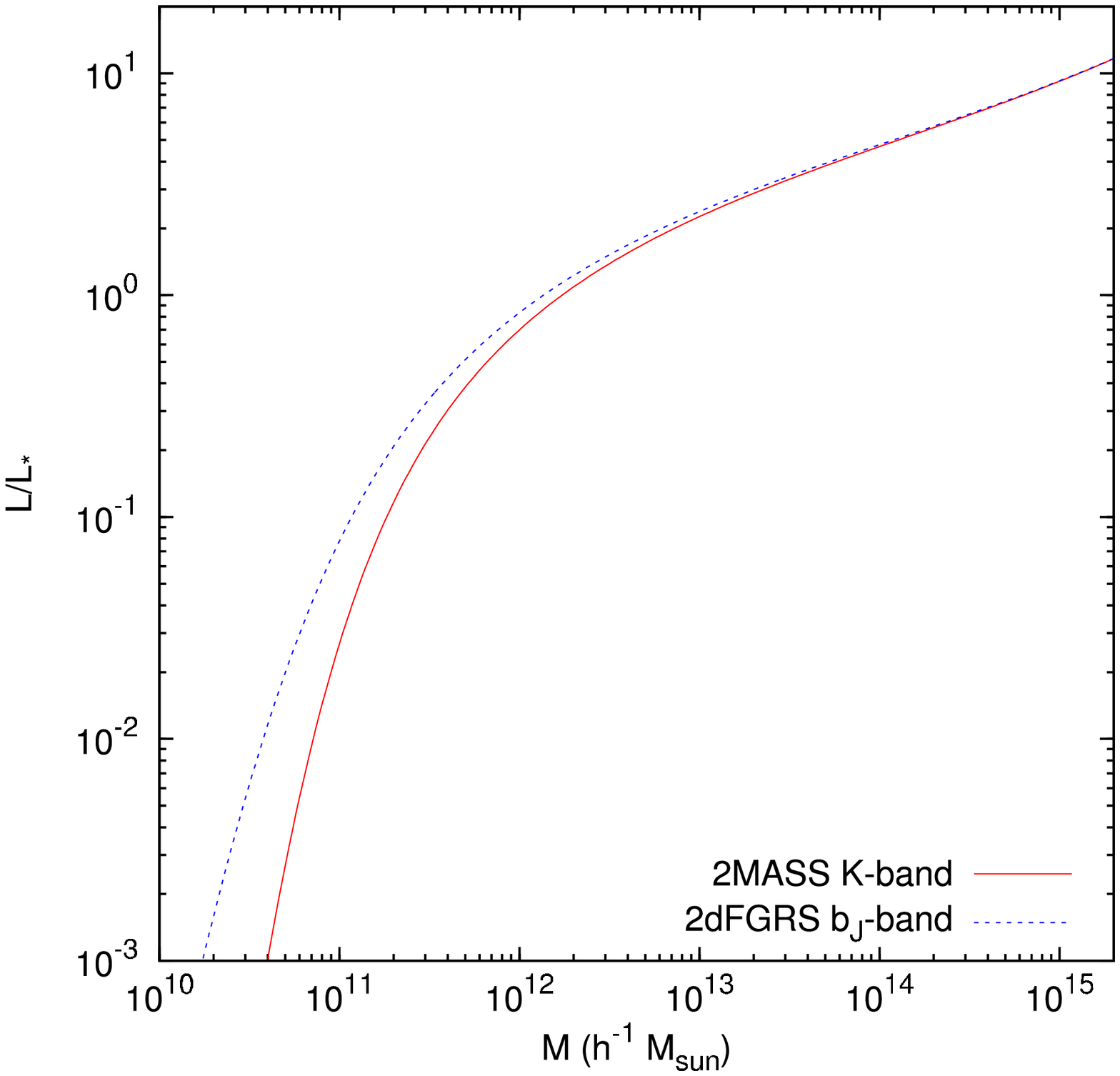}
\includegraphics[width=84mm,angle=270]{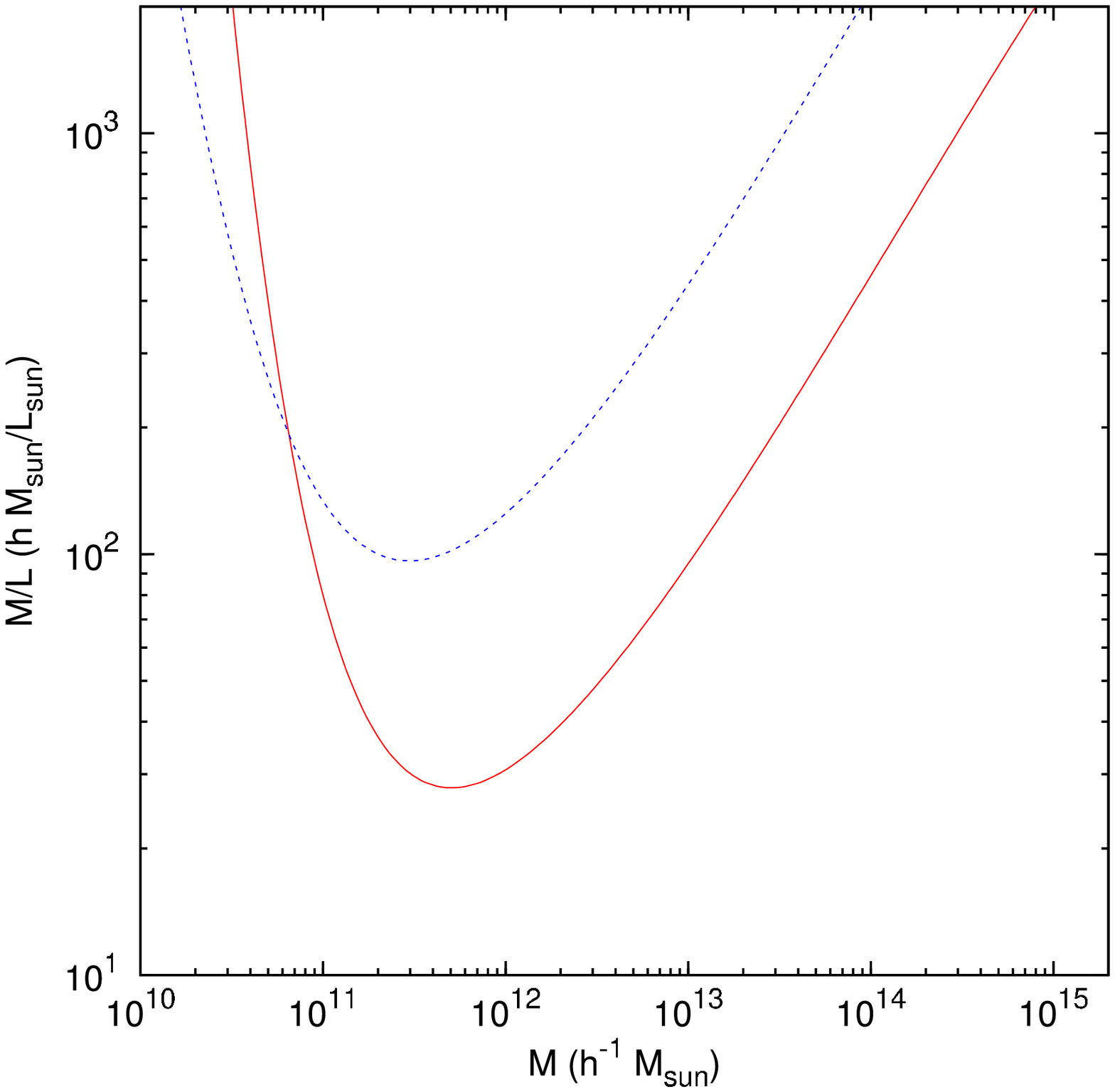}
\caption{Mass-luminosity relation as obtained using the non-parametric model, in
the K- and $b_J-$ bands. Upper pannel shows galaxy luminosity 
normalized to the characteristic luminosity, $L^*$, of each band; lower pannel 
shows the corresponding mass-to-light ratio.}
\label{mlfig}
\end{figure}

Figure \ref{mlfig} shows the results for the luminosity of a single galaxy
as a function of the mass of the hosting halo/subhalo, together with the 
corresponding mass-to-light ratio. Shown are curves for both
K- and $b_J$ bands. We caution that the results for the latter band should be treated
with some reserve. This counting method is not entirely adequate to get 
the mass-luminosity relation in the blue, due to complications arising
from recent star formation, although it is still interesting to compare the 
differences obtained from using two different luminosity functions.

\subsection{Destroyed subhaloes and the subhalo mass fraction}
\label{sect:destroyedsh}

As mentioned, a potentially important correction to the non-parametric model
in paper I is to account for subhaloes which have been completely 
destroyed. The study of this evolution of the 
subhalos and their eventual destruction, with the subsequent merger (or not) of 
their galaxy with the central one, is a very interesting topic by itself, which 
is still not completely understood but which is essential
to a complete understanding of the formation of BCGs. However, such a detailed look
at this question is beyond the scope of the present paper; here, we are merely 
interested in a simple model to account for how much mass was in these destroyed 
subhalos, to correct the normalization of our original subhalo mass function.

Our scheme is based on the fact that most of the luminosity of the central galaxy
is built up by merging with the satellite galaxies brought in by these subhaloes.
In other words, the central (BCG) optical galaxy is made up of the galaxies that have
been ''merged away'' -- disappeared from the original distributions. This is
consistent with what is known of the size, shape and colour properties of 
central galaxies.
We therefore assume that the fraction of mass in these destroyed
subhalos (with respect to the total halo mass), is given by the ratio of the 
central galaxy luminosity to the total luminosity of the halo:

\begin{equation} \label{eq:destfracdef}
f_{dest}=\frac{m_{dest}}{M}=\frac{L_{cent}}{L_{total}} \, .
\end{equation}

\noindent For a given
$L(M)$ relation, which sets the luminosity of both the central and satellite 
galaxies as a function of the halo/subhalo mass, the previous 
equation can then be solved for $f_{dest}$ as a function of halo mass, 
since the total luminosity is
going to be a function of only it and the total mass:
$L_{total}=L_{cent}(M)+(1-f_{dest})L_{sat,max}(M)$, where $L_{sat,max}$ is 
the maximum contribution of the satellites for when $f_{dest}=0$. This is
given by: 

\begin{equation} \label{eq:lsatmax}
L_{sat,max}(M)=\int_0^{0.5 M} L_{ref}(m) N(m|M) dm \, .
\end{equation}

\begin{figure}
\includegraphics[width=84mm,angle=270]{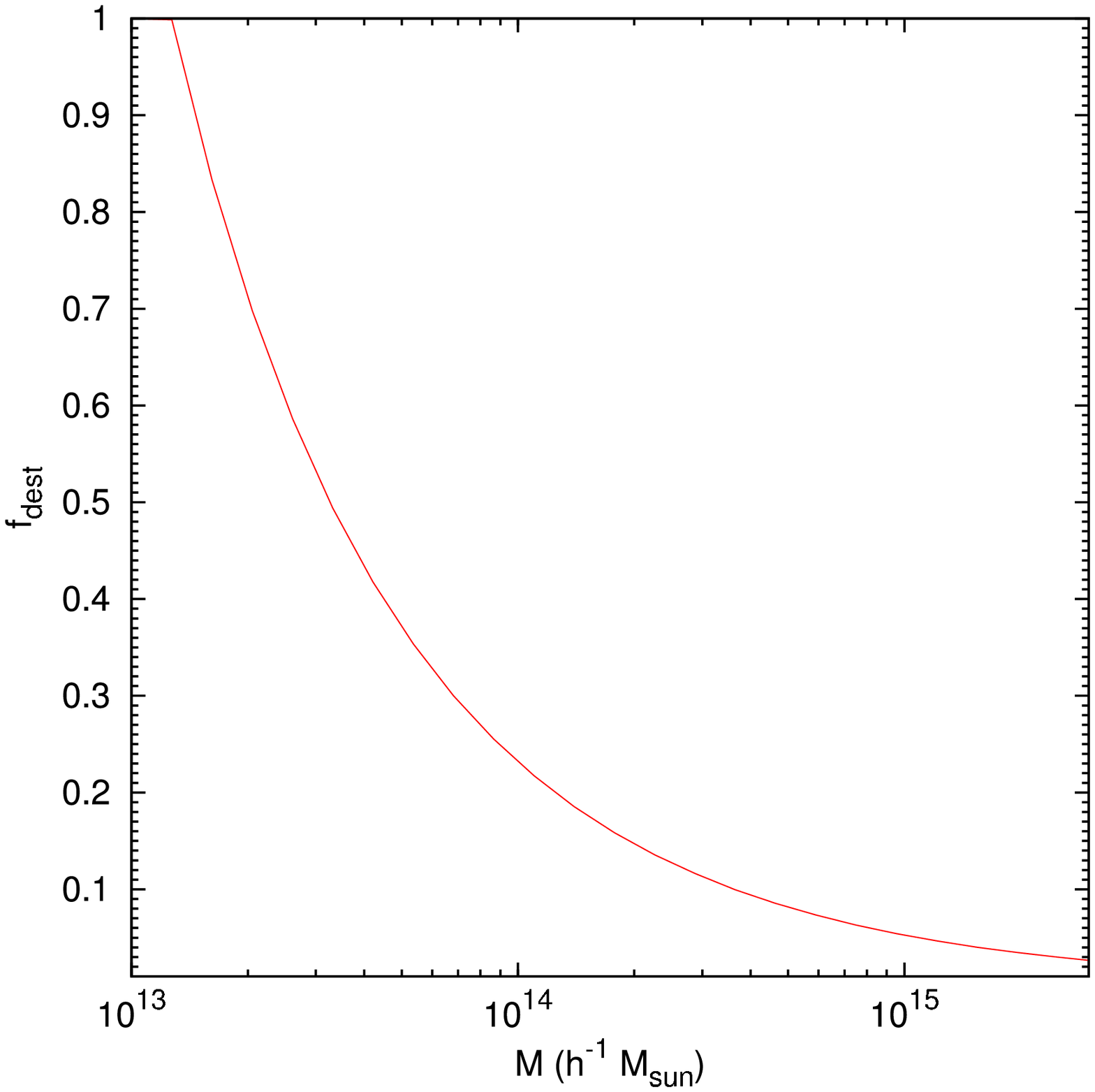}
\includegraphics[width=84mm,angle=270]{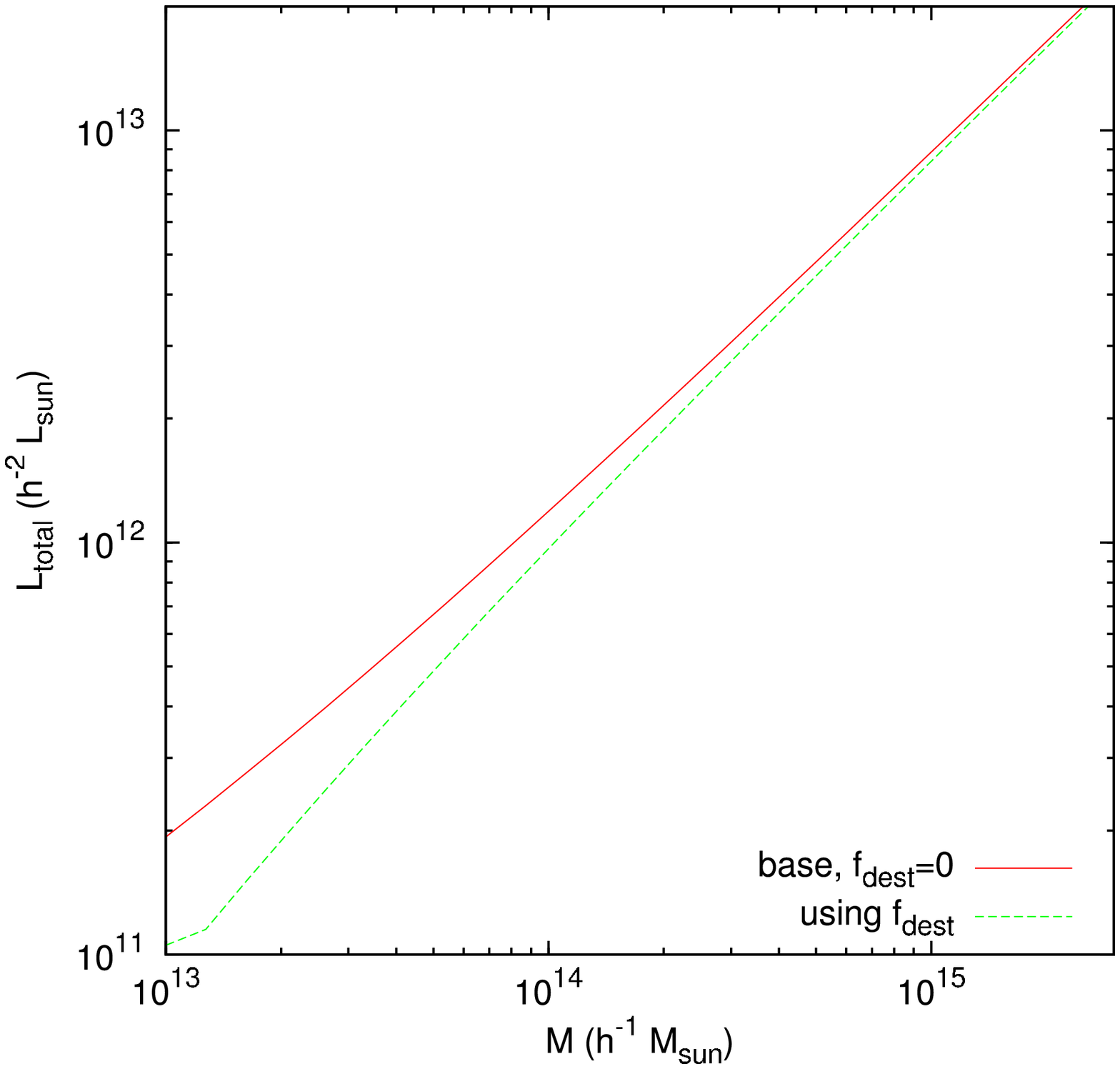}
\caption{Upper pannel: total mass in subhaloes which have been completely disrupted,
as a fraction of the total halo mass.
Bottom pannel: total luminosity as a function of halo mass with or without using the 
fraction of mass in destroyed subhaloes shown in the upper pannel.
Results are for the K-band, using the base mass-luminosity relation of
equation (\ref{mlfit}).}
\label{fig:fdest}
\end{figure}

The upper pannel of figure 
\ref{fig:fdest} shows the destroyed mass fraction as a function of halo mass 
for our base mass-luminosity relation, given by equation (\ref{mlfit}), while
the bottom pannel shows the effect on the total luminosity. 
As can be seen, this is most pronounced at the lower end 
of the mass scale shown, and becomes small enough to have little effect 
at high mass.

There are two additional factors that need to be noted. First, the introduction
of this term can also have an effect on the actual mass-luminosity relation, since
we are using a counting method to obtain it. However, for the mass range we are 
interested in, the number counts are dominated by the central galaxies and will
therefore not be affected (see section 4). Likewise, the only haloes capable
of hosting subhaloes large enough to be counted in this range are the most massive
ones, for which the effect is smallest (see section 3). 
Secondly, in principle, this approach will also depend on the exact form of 
the mass-luminosity relation. However, for the small deviations from the 
base relation we will be considering in this paper, the effect on the total
luminosity is small, since the variation to the base relation will be greatest
at higher mass, where this effect is smallest. We will therefore, for simplicity, 
use this one result throughout the paper.

Finally, it needs to be stressed that this is just a very simple approximation. The 
calculated factor is applied to the whole subhalo mass fraction as a correction
to the normalization, without taking into account a possible dependence on subhalo
mass. In particular, the situation with very low mass subhaloes is very uncertain
in this scheme, since they are expected to be very faint, and have therefore 
very little weight in the sum of the total luminosity, while they can contribute 
an important fraction of the mass. Another important point is that in principle
the luminosity of the galaxies that were contained in the destroyed subhaloes 
should be added to the central galaxy luminosity, since under this scheme we are 
assuming that these are merging. In practice, though, the light in these destroyed
subhaloes is going to be small in comparison with the BCG in this model, since 
the largest fraction of destroyed subhaloes occurs for less massive haloes where 
the BCG is dominant. For simplicity, we will here ignore this contribution.

\section{Introducing scatter}
\label{sect:scatter}

\subsection{The conditional luminosity function}

In the context of the present paper, we need a more detailed
model than the one described previously. Most importantly, it needs to 
include some kind of scatter in the mass-luminosity
relation. Naturally, we expect that not all galaxies in hosts of the
same mass will have the same luminosity. To capture this, we introduce
a dispersion around the average relation describe above. We use the
conditional luminosity function (CLF) formalism introduced by \citet{yang}
(see also \citealt{vdb} and references therein) and by Cooray and
collaborators (e.g., \citealt{cooray,coorayc} and references therein). 
This consists of replacing a deterministic
mass-luminosity relation, like the one in equation (\ref{mlfit}), with
a distribution of luminosity around an average value for any given halo
mass, $\phi_{CLF}(L|M)dL$, which represents the probability of having a 
 galaxy of luminosity $L$ in a halo of mass $M$. Note that here we are only
applying this to the central galaxy in any given halo, since that is the
important one for the study of BCGs; the distribution of satellite galaxies
we draw directly from the distribution of subhalo masses. 

An important point is that this CLF must, by definition, match the
observed luminosity function when it is integrated over all haloes,
i.e. $\int_0^\infty \phi_{CLF}(L|M) n(M) dM=\phi(L)$, where $n(M)$ is
the halo mass function, $\phi(L)$ the observed luminosity function and
$L$ should only be considered in the range where the haloes dominate
the number of hosts (i.e., at high luminosity, which is precisely the
range we are interested in when looking at BCGs;otherwise, we would
also need to account for subhalo contribution).  The introduction of
scatter then leads to a problem with the mass-luminosity relation
derived from the counting method, however. As noted by \citet{iro},
the fact that the mass function is decreasing with increasing mass
causes an effect similar to the Malmquist bias: for any given mass
bin, more objects are scattered into it from lower mass bins than are
scattered out of it. If we then take our base mass-luminosity relation
to be the average one in the CLF distribution, because of this effect
we will end up with a calculated luminosity function that greatly
overestimates the abundance of very bright galaxies when compared to
the observed one.

To get the correct matching to the observed luminosity function, it is then
necessary to modify the average mass-luminosity function we take for the
basis of the CLF. This is achieved by introducing an additional term, 
of the form:

\begin{equation} \label{eq:mlmod}
L(M)=L_{ref}(M)(1+M/M_s)^a \, ,
\end{equation}

\noindent where $L_{ref}(M)$ refers to the base mass-luminosity relation of
equation (\ref{mlfit}). In practice, $a$ is going to be negative since we need
to lower the luminosity corresponding to any given high-mass halo in order
to drive the value of our calculated luminosity function down. 

There is one final, potentially important point about this issue: once scatter is
introduced, care must be taken when looking at the calculated average 
mass-luminosity relation. In our approach, the average luminosity at fixed mass
needs to go down, relative to the scatter-less case or, looking at it the opposite
way, the same average luminosity is obtained for higher mass haloes. This is due 
to the fact that, when considering the CLF, we are doing the binning by mass 
(or more precisely, taking the conditional variable in the distribution to be the 
mass). If we had instead binned 
by luminosity, the effect of introducing scatter would have been the opposite: 
the average mass correspoding to a given luminosity would instead have gone down.
This is to be expected and is just a statistical effect of the two different 
ways in which the conditional function can be defined. It does however mean that
care must be taken when comparing results of different authors to look at how
the binning was done in each case.

In this paper we will consider two different models for the
CLF of the central galaxy: a simpler model where we assume the
distribution is lognormal, but where we are left with a free parameter
in the scatter introduced; and a more complicated one based on the
distribution of concentrations at a given halo mass, which fully
motivates the introduction of scatter in the CLF without any free
parameters. For the satellite galaxies we will use the same modified
mass-luminosity relation as well, since these were central galaxies
within their own independent haloes prior to merging, so it is
reasonable to expect the same effects to apply to them. From
semi-analytical modelling, it has been shown that this is a good
approximation, although a more careful treatment shows a slightly
different relation for sattelites than for central galaxies 
\citep{wang}. But
note that, when doing analytical calculations, using the subhalo 
mass function already introduces a form of distribution for the 
subhaloes as well (in that the mass of a given subhalo can be 
drawn from it, see section 5.1 for further discussion).

\subsection{Lognormal model}

The simplest CLF model we consider is to assume it has a lognormal
form. This is similar to what was done previously by other authors
\citep{cooray,coorayc}, and such a form seems a good
match to the distribution of stellar mass obtained in semi-analytic
modelling \citep{wang}.  The problem with this
approach is that there is no {\it a priori} reason to assume any specific value
for the dispersion. Furthermore, this value is linked to the modified
mass-luminosity relation of equation (\ref{eq:mlmod}), so it needs to
be defined in some way in order to determine the latter. We do this by
determining which value of the dispersion leads to an average
luminosity as a function of mass which best fits observational values.
For simplicity, we will consider that the value of
the dispersion, $\sigma$, is constant and independent of mass. Since
we are only interested in the bright end, where we expect central
galaxies to dominate, and these are known to have only a small scatter
in luminosity (e.g., \citealt{pl,bernardi}), this
is quite likely a good approximation. Semi-analytical modelling also
shows that scatter in stellar mass is only a weak function of halo mass
\citep{wang}.

The luminosity $L$ of a galaxy in a host of mass $M$, is then given by
a lognormal distribution of the type:

\begin{equation} \label{eq:lognormal}
\phi_{CLF}(L|M)dL=\frac{1}{\sqrt{2\pi}\sigma_{LN} L}{\rm exp}
\Big(-\frac{({\rm ln} [L/L_0(M)]+\sigma_{LN}^2/2)^2}{2 \sigma_{LN}^2} \Big) dL \, ,
\end{equation}

\noindent where $L_0(M)$ is some average luminosity for a host of
mass $M$ (discussed further below), and $\sigma$ is the dispersion in
the normal logarithm of the luminosity. Note that there is some
confusion in the literature over the exact form defined for the 
lognormal distribution. First, it is necessary to pay attention
to whether the distribution is in the natural logarithm or base 10 
logarithm of the variable; the quoted value of $\sigma$ will be 
different in the two cases for the same distribution. Secondly,
the way we have defined it in equation (\ref{eq:lognormal}), the 
average luminosity is given by $L_0(M)$. This is due to the 
second term in the denominator of the exponential term, which
not all authors include; if it is omitted, $L_0$ would not be the 
average luminosity.

The difficulty with using an approach such as this is that we are left
with two unknowns we need to determine, the reference luminosity
function $L_0$ and the dispersion $\sigma$. The only condition we can
impose on this distribution is that, when integrated over all hosts,
the resulting luminosity function must match the observed one. Fitting
this calculated luminosity function to the observed one then allows us
to relate the two parameters we have when we take $L_0=L(M)$ from
equation (\ref{eq:mlmod}), $a$ and $M_s$, to the scatter $\sigma$.
However, this still leaves us with one free parameter which we cannot
otherwise specify. In order to address this problem, we determine
which value of the scatter gives us an average luminosity as a
function of mass that best fits the observed data. Since our
original mass-luminosity relation was already quite a good fit 
to the data (see paper I), we must necessarily have only
a small correction to it (i.e., a small value of $a$), which in turn
implies a small value for the scatter, which is qualitatively in
good agreement with observations (see further discussion in section
7).

\subsection{Concentration model}

The other model for the CLF of BCGs we consider is based on
the variation of the concentration of haloes with the
same mass. The basic idea behind this is that the distribution of
concentration in same mass haloes will lead to different mass in the
inner region of the halo where the galaxy will be present; the luminosity
of the hosted galaxy will then simply be proportional to this mass. In 
practice, we calculate the mass-to-light ratio of this inner region, 
for the average concentration and with an average luminosity given by
equation (\ref{eq:mlmod}), and then calculate the change in luminosity
as the concentration changes by assuming that the mass-to-light ratio
is fixed (observationally, it has been noted that the dynamical mass-to-light
ratio of BCGs is almost constant, e.g. \citealt{linden}).
This then gives us the BCG luminosity as a function of both concentration
and halo mass. 

Although the model is conceptually simple, the details are problematic.
The main issue is to determine what exactly is this inner region
and how to calculate its mass. The most obvious solution, taking
quoted values from the literature for BCG radius and its dependence
on luminosity, is not really satisfactory since these are most often
determined from isophotal limits and in the case of cD galaxies it would
be necessary to further consider whether to include the envelopes;
also, it is natural to assume that the actual region of influence for 
the dark matter is more extensive than the visible galaxy. Other
options such as some parameter from the dark matter structure, 
run into the problem of motivating what exactly it should be.

In the end, after checking the results of several different possible 
models, we concluded that the one that has the best motivation and 
also gives the best results is to calculate the inner region mass
by using a weighting function based on the luminosity profile of 
the BCGs. 

Since we just require it to make our weighting function, for
simplicity we assume that the luminosity profile of BCGs
can be universally fit by a Sersic profile:

\begin{equation} \label{eq:sersic}
I(r)= A\, exp(-b_n[(r/r_e)^{1/n}-1])\, ,
\end{equation}

\noindent where $A$ is a normalization factor, and $\Gamma(2n)=2\gamma(2n,b_n)$,
where $\Gamma(2n)$ is the gamma function and $\gamma(2n,x)$ 
the incomplete gamma function; this can be well approximated by
$b_n\approx 2n-0.327$ \citep{capaccioli}. 
It is know that there is a correlation between the
profile parameters $n$ and $r_e$, and also between $r_e$ and
the galaxy luminosity $L$, although the correlation between
$n$ and $L$ is very weak (e.g., \citealt{graham}). For simplicity, 
we will assume that we can relate both parameters to the
galaxy luminosity; although in practice this is not really true,
for our purposes here it is a sufficient, if rough, approximation.
Based on results from the literature \citep{graham,lm}, we use the
following relations:

\begin{equation} \label{eq:nsersic}
n=2.8694{\rm log}(r_e)+2.0661 \, ,
\end{equation}

\begin{equation} \label{eq:resersic}
{\rm log}(r_e)=0.9523{\rm log}(L)-9.1447 \, ,
\end{equation}

\noindent where $L$ is the K-band luminosity; these are similar to what
is reported by other authors (e.g., \citealt{bernardi}). Our weighting function
is then given not by the actual profile, but rather the integration
factor for the luminosity, $w(r)=r I(r)$, and the normalization
$A$ chosen so that $\int_0^{r_{vir}}w(r)dr=1$. Finally, the inner region
mass is simply obtained by integrating the mass density times the
weighting function:

\begin{equation} \label{eq:innermass}
M_{inner}=\int_0^{r_{vir}}4\pi r^2 w(r) \rho(r) dr \, .
\end{equation}

We use the usual NFW profile for the density \citep{nfw},

\begin{equation} \label{nfw}
\rho=\frac{\rho_1}{x(x+1)^2} \, ,
\end{equation}

\noindent where $x=r/r_s$, with $r_s=r_{vir}/c_{vir}$ and the virial
radius is $r_{vir}=(3 M/4 \pi \Delta_{vir} \bar{\rho})^{1/3}$ and
$\Delta_{vir}=387$; $\rho_1$ is normalized to give the halo mass at
the virial radius.

For the concentration distribution, we take the model of
\citet{bullock} (see also \citealt{maccio}, who get similar results). 
This relates the average concentration of a halo of
mass $M$ with the scale factor at its collapse, $a_c$, given by:

\begin{equation} \label{acoll}
\sigma(f M)=\frac{\delta_c}{D(a_c)} \, ,
\end{equation}

\noindent where $\sigma$ is, as usual, the variance of the linearly
extrapolated power spectrum of perturbations, $D(a)$ is the growth
factor and $\delta_c=1.673$ the linear threshold for collapse;
$f=0.001$ is a parameter. The concentration is then given by
$c_{vir}=k/a_c$, with the parameter $k=3$. Finally, the distribution
of the concentration is given by a lognormal distribution with average
$c_{vir}$ and variance $\sigma[{\rm log}(c_{vir})]\sim
0.18$. 

The BCG luminosity distribution is then obtained from the concentration
distribution by $\phi(L|M)=f(c|M)/(dL/dc)$, where $f(c|M)$ is the 
concentration distribution as a function of halo mass. As mentioned,
the luminosity for any given concentration and halo mass is given by

\begin{equation} \label{eq:lumconc}
L(c,M)=\frac{M_{inner}}{(M/L)_0} \, ,
\end{equation}

\noindent where $(M/L)_0$ is the mass-to-light ratio with 
the inner mass calculated at the average concentration and the 
luminosity given by our mass-luminosity relation, equation (\ref{eq:mlmod}),
and $M_{inner}$ is given by equation (\ref{eq:innermass}).
Finally, we fit our calculated luminosity function to the observed
one, in order to determine the parameters that go into the 
modified mass-luminosity relation (see table 2). 

Note that both \citet{bullock} and \citet{maccio} find that subhaloes
tend to have higher concentrations than parent haloes of the same
mass. Although it goes beyond the scope of the present work, it is
wortwhile to mention that in the framework of the model just presented,
this can possibly lead to slightly different distributions of
luminosity as function of mass for the subhaloes, although this is
most likely complicated by the fact that we need to take the subhalo
properties at accretion rather than at present.

\begin{table*} \label{sigmamodparam}
\begin{center}
\begin{tabular}[c]{|c|c|c|c|}
model & $\sigma_{LN}$ & $a$ & $M_s$ \\
\hline
concentration & N/A & -0.08 & $10^{13.5}$\\
lognormal & 0.265 & -0.07 & $10^{13}$\\
\end{tabular}
\caption{Fit parameters for the modified mass-luminosity relation
of equation (\ref{eq:mlmod}), for the concentration and
lognormal models. The dispersion in the latter is fixed to the value 
shown, and variable in the former.}
\end{center}
\end{table*}

\begin{figure}
\includegraphics[width=84mm,angle=270]{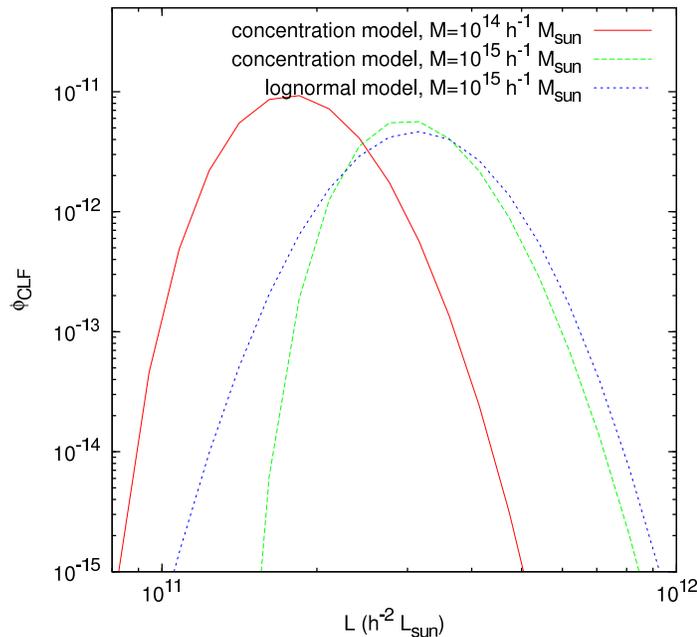}
\caption{Examples of the calculated distribution for the BCG luminosity,
for both the lognormal and concentration models, and for two different values
of the halo mass.}
\label{fig:clfdistribution}
\end{figure}

Table 2 shows the values we obtain for the parameters of the 
modified mass-luminosity relation of equation (\ref{eq:mlmod}). 
Figure \ref{fig:clfdistribution} shows examples of the actual
distribution we obtain for the BCG luminosity, both for the lognormal
model and for the concentration model.

\section{Central vs satellite galaxies}

As was briefly mentioned above, the way we build up the
mass-luminosity relation naturally gives rise to a model for clusters,
featuring a distinct separation between central and satellite
galaxies. This comes from the fact that we consider that galaxies are
hosted by both the parent halo and the subhaloes. Since we consider
that the same mass-luminosity relation applies for both, and the
former will be, by definition, considerably more massive than the
latter, this results in there being a central, very luminous galaxy,
hosted by the parent halo, while fainter satellite galaxies are spread
throughout in the subhaloes. 

Of particular importance for the question of whether the first
brightest galaxies are special, this separation implies that
these galaxies should indeed have a special luminosity 
distribution, independent from
that of the remaining galaxies in the cluster.  This comes from the
fact that the distribution functions of these two types of galaxies
will be different in origin: the central galaxies one will be
determined by the halo mass function, while the satellite galaxies one
will depend on the subhalo mass function.  This dichotomy is also
found in HOD models, for instance when accounting for total galaxy
occupation number (e.g., \citealt{yanghod,zz}): while $P(N|M)$, the
probability that a halo of mass $M$ hosts $N$ galaxies, is Poissonian
at high $N$, where satellite galaxy numbers dominate, it is
significantly sub-Poissonian at low $N$, indicating that the
distribution of the central galaxies is much more deterministic.

This has an important consequence, derived from the fact that, at high
mass, and therefore also at high luminosity, the total mass function
is dominated by the haloes, not subhaloes. This means that, when
analysing the luminosity function of galaxies in clusters, the
brightest region will be dominated by the central galaxies, which will
actually be more abundant overall than the brightest of the satellite
galaxies. We then expect that this
will cause a feature in the cluster galaxy luminosity function at
the bright end; this point will be examined in further detail
below. Another consequence is that we expect the luminosity of
the central galaxy in the cluster to be completely determined by halo
mass and its distribution, without the need to account for surviving 
subhaloes.

\begin{figure}
\includegraphics[width=84mm,angle=270]{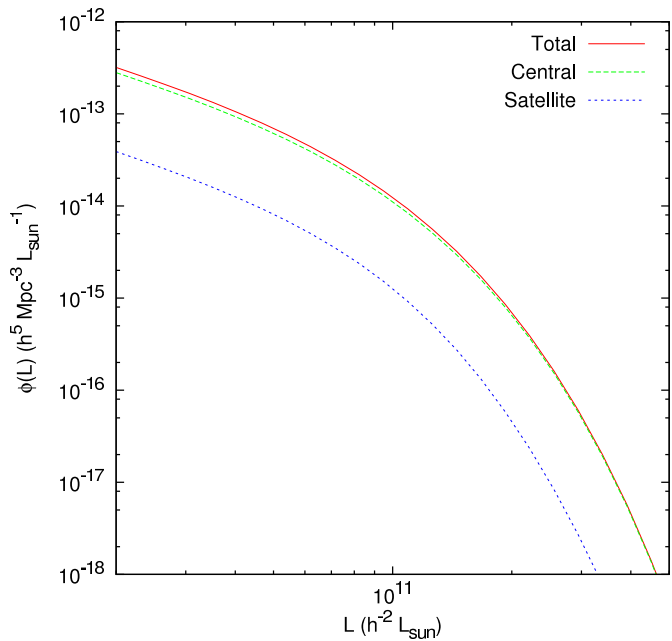}
\caption{Contribution to the high-end luminosity function of central and
satellite galaxies. These particular curves are for the $K$-band,
lognormal model, but the results are similar for the concentration 
model. It is very noticeable that the halo numbers dominate in this
luminosity range. The relative satellite contribution to the total luminosity
function is qualitatively similar to what is found by other authors
\citep{vdb}.}
\label{centrallffig}
\end{figure}

It is in fact possible to derive the different contributions to the
global luminosity function from the central and satellite galaxies, by
associating them with the halo and subhalo distributions,
respectively, and then using the CLF formalism presented in the
previous section:

\begin{equation} \label{xform}
\phi_i(L)dL=\int_0^\infty \phi_{CLF}(L|M) n_i(M)dM \, ,
\end{equation}

\noindent where the $i$ indexes refer to the haloes and subhaloes,
respectively. The derived luminosity functions for central and
satellite galaxies are shown in figure
\ref{centrallffig}. Unsurprisingly, the central galaxies completely dominate the
overall luminosity function at high luminosity, with their numbers
becoming comparable to the satellites only at low luminosity. This simply
reflects the trends seen in the halo and subhalo numbers (see paper
I).  The expected relative contributions of both types of galaxies are
still uncertain: while \citet{bfb} using their semi-analytical
modelling find satellite galaxies to dominate at the faint end,
\citet{coorayb} using a conditional luminosity function formalism find
that central galaxies dominate throughout the range (likewise,
\citet{zz} find that central galaxies dominate the stellar mass
function on any mass scale).

\subsection{Definition of cluster threshold mass} 
\label{sect:clmass}

A necessary first step before continuing this analysis is to define
precisely what we mean by a ''cluster''. We will opt for a simple choice,
following the standard Abell definition of rich cluster, namely that
it must have upwards of 30 objects brighter than $m_3+2^m$, where
$m_3$ is the magnitude of the third brightest galaxy in the
cluster. It is then possible, following our model, to translate this
into a minimum mass threshold for a halo to host a cluster, as
follows.

The third brightest galaxy will correspond to the second most massive
subhalo (since the brightest galaxy is hosted by the parent halo
itself), and the probability of this having a mass $m_2$ is then:

\begin{equation} \label{2ndmass}
P_2(m_{s,2},M_h)=N(m_{s,2}|M_h) <N> {\rm e}^{-<N>} \, ,
\end{equation}

\noindent where $N(m_{s,2}|M_h)$ is the mass distribution function of the
subhaloes, equation (\ref{shmf}), and $<N>=\int_{m_{s,2}}^\infty N(m'|M_h)
dm'$ is the average number of subhaloes more massive than $m$ in a
parent halo of mass $M_h$. This expression assumes that the
distribution of subhalo masses is Poissonian with average $<N>$, as
expected for the subhaloes (e.g., \citealt{kravtsov}; since we are
looking at cluster sized haloes, $<N>$ will be large in this case),
and it is simply the product of the probability of having a subhalo
with mass $m_2$, given by the first term on the right hand side, by
the Poisson probability of having exactly one subhalo more massive
than $m_2$. Using the fact that $d<N>/dm=-N(m_s|M_h)$, it is easy to
check that this probability is well normalized to 1. The average value
of the magnitude corresponding to this subhalo, $m_3$, can then be 
calculated from the distribution by:

\begin{equation} \label{2ndmassavg}
<m_{3}(M_h)>=\int_0^\infty m(m_s) P_2(m_s,M_h) dm_s \, ,
\end{equation}

\noindent where $m(m_s)$ represents the corresponding magnitude as a
function of the subhalo mass, calculated using the mass-luminosity
relation from section 2. In this instance, we have used the simpler
relation of equation (\ref{mlfit}), since it greatly simplifies the
calculations and using the full CLF results in only a slight
difference.

\begin{figure}
\includegraphics[width=84mm,angle=270]{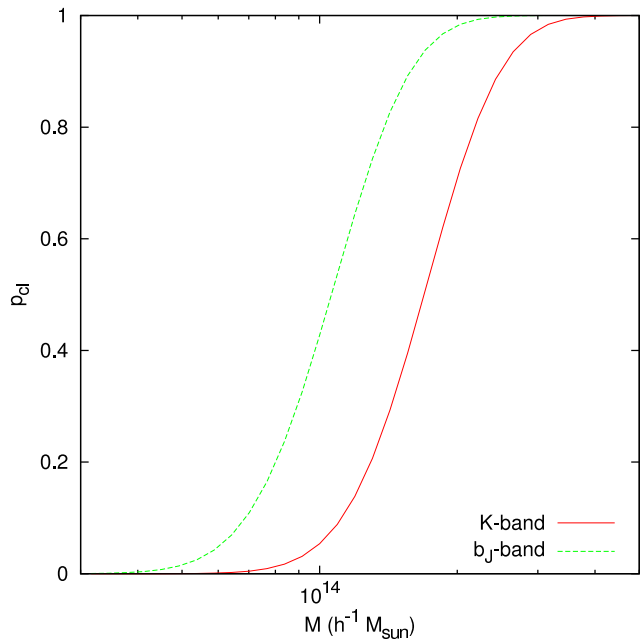}
\caption{Probability that a halo of mass $M$ contains more than 30
galaxies brighter than $m_3+2^m$ and is considered a rich cluster
according to the usual Abell definition. Using either the $b_J$ or the $K$
bands to do the counting results in the two different curves.}
\label{clustprob}
\end{figure}

Using this magnitude we can then obtain $m_3+2^m$, and then convert
this back into a mass threshold, $m_t(M_h)$, which will be dependent
on the parent halo mass.  Finally, we can find the probability, as a
function of $M_h$, that $N(m_t|M_h)\geq 29$ (giving more than 30
objects above the magnitude limit, when including the central
galaxy). Since we are assuming that the subhalo distribution is
Poissonian, this will simply be given by $\sum _{n=29}^{\infty}
P(n,\mu )$, where $P(n,\mu )$ is the normal Poisson probability with
average $\mu=<N(m_t|M_h)>$. This gives a smooth transition for the
mass of cluster hosting haloes, shown in figure \ref{clustprob},
starting around a halo mass of $M_h=10^{14} h^{-1} {\rm
M_\odot}$, but which depends on the luminosity function considered.

\subsection{Cluster galaxy luminosity function}

\begin{figure}
\includegraphics[width=84mm,angle=270]{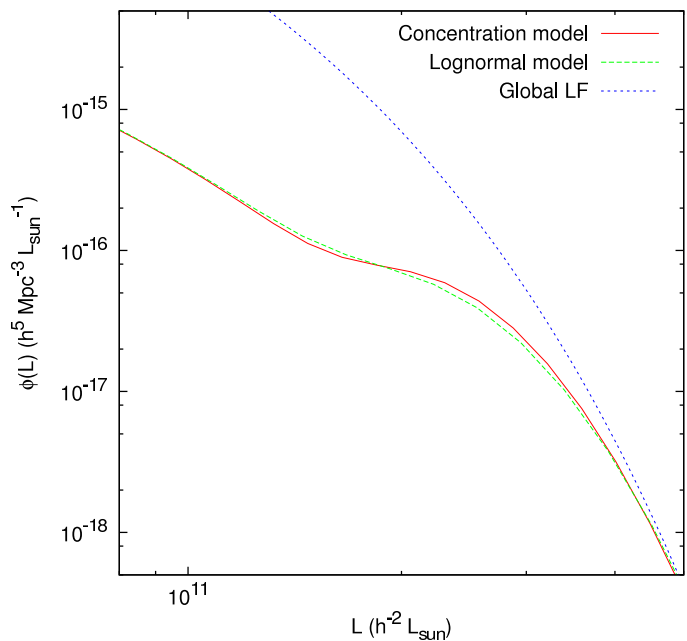}
\caption{Luminosity function of galaxies in rich clusters, in the $K$-band. 
The two curves correspond to the two models for the dispersion used,
described in the previous section. Also shown, for comparison,
is the global LF. Only galaxies above the mass threshold shown in 
figure \ref{clustprob} were considered.}
\label{cglffig}
\end{figure}

Once we have a mass threshold for haloes hosting clusters, obtaining
the cluster galaxy luminosity function is straightforward: we simply
sum up the mass function of haloes above this threshold with the mass
function of all the subhaloes hosted by them (i.e., use equation
(\ref{shmf}), but further multiplied by a term to reflect the
probability that the halo does indeed host a rich cluster, given by
the result shown in figure \ref{clustprob}). Then, we transform this
into a luminosity function using the CLF. 
Our result is shown in figure \ref{cglffig}.

Qualitatively, we obtain a good agreement with observed luminosity
functions (such as the one of \citealt{2dfcluster}, although in this
particular case a direct comparison is difficult because these results
are in a different band; redoing our analysis in the same band
produces a good match), particularly in the lower luminosity range. At
the bright end, there is some disagreement caused by a particular
feature of our result, a bump in the luminosity function at the bright
end. It is simple to understand that this bump is caused by the
central galaxies. The discrepancy in numbers between central and
satellite galaxies comes from the fact that, at high luminosity, the
contribution from parent haloes to the total luminosity distribution
(shown in figure \ref{centrallffig}) completely dominates over the
subhalo one. This is a reflection of the fact that haloes are much
more abundant at the high mass end than subhaloes (see paper I). In
fact, it can be seen from the figure that these central galaxies
essentially correspond to the high luminosity end of the global
luminosity function. This is hardly surprising, since we can expect
the most luminous galaxies to lie at the centre of the most massive
clusters.

Such a feature is thus a natural consequence of the model: very
luminous galaxies will predominantely be central galaxies of high mass
haloes, which will therefore dominate in number over satellite
galaxies of the same luminosity; at the same time, since we introduce
a lower mass limit to rich cluster hosting haloes, the faint end
contribution to the luminosity function of galaxies in such clusters
will come entirely from satellites. It is important to note that this
is not a particularity of this specific model: any model associating
central galaxies with parent haloes and satellite galaxies with
subhaloes will show a similar feature, due to the discrepancy in
numbers between the two at high mass (though it may also require that
this be associated with high luminosity in both cases; or, more
particularly, that the same mass-luminosity relation is used for both
haloes and subhaloes, as is the case in the model used here). 

A note of caution comes from the fact that the shape of the bump
depends on the dispersion in the CLF: the smaller it is, the sharper
the bump will be. Furthermore, the actual shape of the bump is determined
by the cluster definition being used, through the cluster threshold
mass discussed in the previous section. This cutoff mass is
responsible for the decreasing values of the cluster galaxy luminosity
function on the left side of the bump; a lower threshold mass would
result in a wider bump. Taking this threshold to lower and lower
values (beyond the range where it would be reasonable to assume the
presence of clusters) results in the progressive disappearance of the
bump as we naturally regain the overall global luminosity function. It should
be stressed, however, that the presence of a bump is a fundamental
prediction of the model, independent of the precise cluster definition
being used, since it is a direct consequence of the discrepancy in
numbers between haloes and subhaloes at high mass.

This kind of feature is also present in some recent work dealing with
HOD models and the central/satellite galaxy separation. In \citet{zz}
(see also \citealt{zhu}),
the authors use a semi-analytical model of galaxy formation to obtain
the conditional galaxy baryonic mass function. For high mass haloes in
the range we are considering here, they also obtain a high mass bump
in this function caused by the central galaxy. They show that the
baryonic mass function can be described by combining a Schechter
function representing the satellite galaxies contribution with a high
mass gaussian due to the central galaxy. Likewise, \citet{zehavi}
build up HOD models from SDSS results, and analyze the central/satellite 
galaxy split; from this, they build conditional luminosity
functions, and show that their results imply that the central galaxies
lie far above a Schechter function extrapolation of the satellite
population. The observational work also hints at similar features: for
example, it is known that cD galaxies are brighter than what is given
by the bright end of the cluster galaxy luminosity function (e.g.,
\citealt{cdreview}). This is also present in studies of the luminosity
function of galaxies in clusters (e.g., \citealt{2pigg}).  All this
once again reinforces the notion that central galaxies form a special
distribution, essentially separate from the satellite galaxy one.

\section{Building the BCG luminosity distribution}

As discussed above, the non-parametric model used naturally builds a
picture of galaxy clusters. This translates itself into a procedure to
build up the total luminosity distribution of galaxies within a halo of a given
mass. This will be composed of two steps, one dealing with the
satellite galaxies in the subhaloes, another with the central galaxy.

\subsection{Satellites}

In this step, we need to sum up the total luminosity in the satellite
galaxies contained in the halo. We start by taking the total number of
subhaloes in a given halo, as calculated from the SHMF 
(that is, the occupation number; see paper I)
, as an average number for a parent halo of this
mass, taking into account the effect of destroyed subhaloes as 
introduced in section 2.1. 
In this step it is necessary to specify a minimum mass: we take
a low enough value to ensure that we account for all subhaloes massive
enough to give a noticeable contribution to the total luminosity of
the halo. We then assume that the total number of subhaloes follows a
Poisson distribution (as discussed above; e.g., \citealt{kravtsov}).

For each subhalo, up to a total as calculated from the Poisson
distribution, we determine a mass, by assuming the subhaloes follow a
random distribution given by the subhalo mass function
(\ref{shmf}). Then we convert this to the luminosity of the hosted
galaxy using the mass luminosity relation.

Finally, once we have the total number of subhaloes (as determined
initially from the Poisson distribution), we can sum all of their
calculated luminosities to obtain the total in satellite galaxies.

At the same time, the average luminosity of the brightest satellite
galaxy can be calculated in a more direct fashion. Using the subhalo
mass distribution, we can get the probability distribution of the mass
of the most massive subhalo. Analogously to what was done in the
previous section for the second most massive subhalo (see equation
\ref{2ndmass}), this will be given by

\begin{equation} \label{1stmass}
P_1(m_1,M_h)=N(m_1|M_h) {\rm e}^{-<N>} \, ,
\end{equation}

\noindent where $<N>=\int_m^\infty N(m'|M_h) dm'$ as before. Used
together with the mass luminosity relation, the average luminosity of
the galaxy hosted in the most massive subhalo (and therefore the most
luminous of the satellites) is simply given by
$<L_1>(M_h)=\int_0^\infty L(m_1) P_1(m_1,M_h) dm_1$.

\subsection{Central galaxy} \label{centbuild}

The way we have built up the CLF gives us a natural way of obtaining the
distribution of the luminosity of first brightest galaxies with
cluster luminosity, since we are already introducing a distribution
with mass. We use the results of both the lognormal and concentration
models for comparison.

The total luminosity is obtained by summing over the BCG and satellite
contributions, and taking into account the effect of the destroyed
subhaloes from section 2.1. The average total
luminosity at any given mass is simply the sum of the average
luminosities of the BCG and satellites, and is, by construction, equal
to the one shown in figure \ref{fig:fdest}.

Finally, we can also obtain the global distribution of BCG luminosity
over all clusters. To do this, we use the cluster threshold mass, as
calculated in the previous section, and simply integrate the
conditional distribution we have multiplied by the halo mass function:

\begin{equation} \label{bcgglobal}
f(L_1)=\int_0^\infty f(L_1|M) n(M) p(M) dM \, ,
\end{equation}

\noindent where $n(M)$ is the halo mass function, given by
(\ref{stmf}), and $p(M)$ is the probability that a halo of mass $M$
hosts a rich cluster, as given in figure \ref{clustprob}.

\section{Other models for BCGs}

In this section, we introduce two other simple models to complement
the one presented above in order to allow better comparisons with the
observational results. The first, and, a priori, best motivated is
based on the simple assumption that the BCGs are merely the extreme
values of the unique distribution that applies to all cluster
galaxies. We do this based on a regular cluster galaxy luminosity
function, and build the BCG distribution directly from it. The second
is based on assuming some form of cannibalism. We use a simple
approach to model this mechanism, that of merging the brightest galaxy
with one other, where both are taken from the universal distribution
in the first case we consider.

\subsection{Extremes of a general distribution} \label{xtremes}

The simplest assumption possible when studying the distribution of the
galaxies in a cluster is that they are all drawn from the same
statistical distribution (for example, a Schechter function for
luminosity). The galaxies in a cluster would then simply be a random sample
drawn from this distribution, with the brightest galaxy simply the
extreme value of this sample. This approach has been studied in the
literature before (e.g., \citealt{tr}). In particular, it has been shown
that this approach leads to a result from extreme value theory, the
Gumbel distribution, for the overall distribution of the magnitudes of
the first brightest galaxy \citep{bb84,bb}. This is given by 

\begin{equation} \label{gumbel}
f(M)=a e^{a(M-M^*)-e^{a(M-M^*)}} \, ,
\end{equation}

\noindent with $M^*=M_G+\frac{0.577}{a}$, where $M_G$ is the mean of
the magnitude values and $a$ is a measure of the steepness of fall of
the parent distribution.

In the present paper, we take a slightly different approach, in order
to match that which we will take for the other models. This model and
subsquent calculations are similar to the ones presented in
\citet{tr}, although some details, like the luminosity function used,
will be different.  We begin by assuming that the parent distribution
of the luminosity of the galaxies in a cluster is given by a Schechter
function, equation (\ref{schechter}) (for simplicity, we use the same
values for $M_*$ and the faint end slope, $\alpha$, 
as the global luminosity function).

The only dependence on the actual cluster considered comes
in the normalization, which we set so that the total luminosity in all
the galaxies equals the cluster luminosity, $L_c$:

\begin{equation} \label{clusternorm}
\phi_*(L_c)=\frac{L_c}{\Gamma[2-\alpha] L_*} \, .
\end{equation}

We then take the value of the distribution of the first brightest
galaxy at a given luminosity to be the probability that there are no
galaxies brighter than that luminosity, times the probability that
there is a galaxy at that luminosity. The latter is simply given by
(\ref{schechter}), while for the former we take a Poisson fluctuation
around the average number of galaxies obtained by integrating the
parent distribution, (\ref{schechter}). Thus, the luminosity
distribution for the brightest cluster galaxy is given by:

\begin{equation} \label{dist_1bg}
f_1(L,L_c)=\phi(L,L_c) e^{-\int_L^\infty \phi(L,L_c)dL} \, ,
\end{equation}

\noindent where the integral in the exponential can be resolved to
$\int_L^\infty \phi(L)dL=\phi_*(L_c) \Gamma[1-\alpha,L/L_*]$. Using
the same principles as discussed in section \ref{sect:clmass} above, it is
simple to show that this probability distribution is
adequately normalized to 1.

Similarly, the probability for the luminosity of the second brightest
galaxy is given by the product of the probability of having a galaxy
at a given luminosity by the probability of there being a single
galaxy brighter than that luminosity. In general, and again taking Poisson
fluctuations around the average number, the distribution of the n-th
brightest galaxy will be given by:

\begin{equation} \label{dist_n}
f_n(L,L_c)=\phi(L,L_c) \big(\int_L^\infty \phi(L')dL'\big)^n e^{-\int_L^\infty
\phi(L')dL'}/n! \, .
\end{equation}

Likewise, the joint probability of having the first brightest galaxy at
luminosity $L_1$ and the second at $L_2$ (given by the product of the
probability of a galaxy at $L_2$, another at $L_1$, none in between
and none above $L_1$) can also be derived:

\begin{equation} \label{dist_12bg}
f_{12}(L_1,L_2,L_c)=\phi(L_1,L_c) \phi(L_2,L_c)
e^{-\int_{L_2}^\infty \phi(L)dL} \, .
\end{equation}

Once we have these distributions, it is then easy to obtain the
various statistics we are interested in: the average magnitude of the
first and second brightest galaxies, $<m_1>$ and $<m_2>$, the
dispersion in first brightest galaxy magnitude, $\sigma_1$, and the
average magnitude difference between these two
$<\Delta_{12}>=<m_1-m_2>$. Some caution is necessary with the last
one, since the fact that $L_1$ and $L_2$ are not independent variables
would necessitate the use of the joint distribution; however, it turns
out that the integrals over this joint distribution resolve to two
different integrals over the two separate distributions, so that
$<\Delta_{12}>=<m_1>-<m_2>$ as calculated for each separate,
individual distribution.

\subsection{Cannibalism}

To illustrate the effect of galactic cannibalism, we consider here two
simple models, starting with a common distribution like the one
discussed in the previous section, and then merging the brightest
galaxy with one of the others.

In general, the distribution of the new BCG will have to be
taken from the joint distribution of the previous first and n-th
brightest galaxies. It is important to note that these variables are
not independent, and as such when calculating the average of the sum
it will not, in principle, be possible to separate it into integrals
over the two individual distributions (except in the particular case
when $n=2$, as noted above).

\subsubsection{$L_1+L_2$}

The simplest possible model to consider when taking account of
cannibalism is to merge the two brightest galaxies from a common
distribution. Thus, the luminosity of the brightest galaxy is now
given by the sum of the luminoisities of the previous two brightest
galaxies, with a distribution given by the joint distribution of
equation (\ref{dist_12bg}). The second brightest galaxy is now the old
third brightest, with a distribution given by equation (\ref{dist_n}), 
with $n=3$.

The remaining calculations follow the same pattern as discussed in the
previous case. It is to be expected that, due to the fact that the new
first brightest luminosity is the result of the sum of two previous
variables, the dispersion in first brightest luminosity will be
slightly lower than before. Likewise, it is obvious that the
difference in magnitude between first and second brightest galaxies
will now be much larger.

\subsubsection{$L_1+L_n$}

A slightly more realistic toy model to illustrate galactic cannibalism
is to proceed similarly as described above, but instead of merging the
first brightest galaxy with the second, to take some weighting function
to reflect the probability that the merger will occur with any one
other of the galaxies in the cluster.

We take the probability of the merger occuring with the n-th brightest
galaxy to be proportional to its average luminosity, $L_n$; this gives
greater weight to the first few brightest galaxies, but since the
average luminosity decreases only slowly with $n$, the probability is
split over a wide range. We consider a possible merger down to the
30th brightest galaxy in the cluster, for which the merger probability
given by this prescription will be below 1\%.

The total probability distribution of the new BCG luminosity is
obtained from the joint probability distribution of the different
weighted pairs. Since the variables are not independent and therefore
this joint distribution cannot be split into a product of terms each
dependent only on a single variable, to calculate $<m_1>$, $<m_2>$ and
$\sigma_1$ it is necessary to solve complicated integrations. We turn
instead to a Monte Carlo method, building up the distribution of the
new BCG luminosity by randomly generating merger pairs and the
luminosity of their components, and then summing them. The new second
brightest galaxy will be either the old one, or the old third
brightest, depending on whether the merger occurred with the former or
not.

\section{Results}

\begin{figure}
\begin{center}
\includegraphics[width=84mm,angle=270]{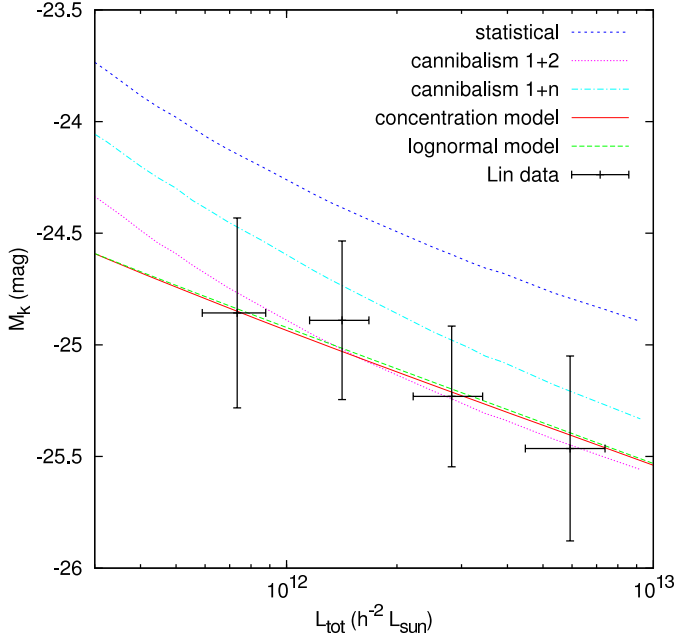}
\caption{Average magnitudes, in the $K$-band, of the first brightest
galaxy, as a function of the total cluster luminosity, as calculated
for the different models: the ones based on the CLF formalism
introduced, the statistical model and two forms of cannibalism: the
extreme $L_1+L_2$, and the softer $L_1+L_n$.  The data points shown
are binned values of cluster galaxy data supplied by Yen-Ting Lin
(2006, private communication).}
\label{m1avgfig}
\end{center}
\end{figure}

\begin{figure}
\begin{center}
\includegraphics[width=84mm,angle=270]{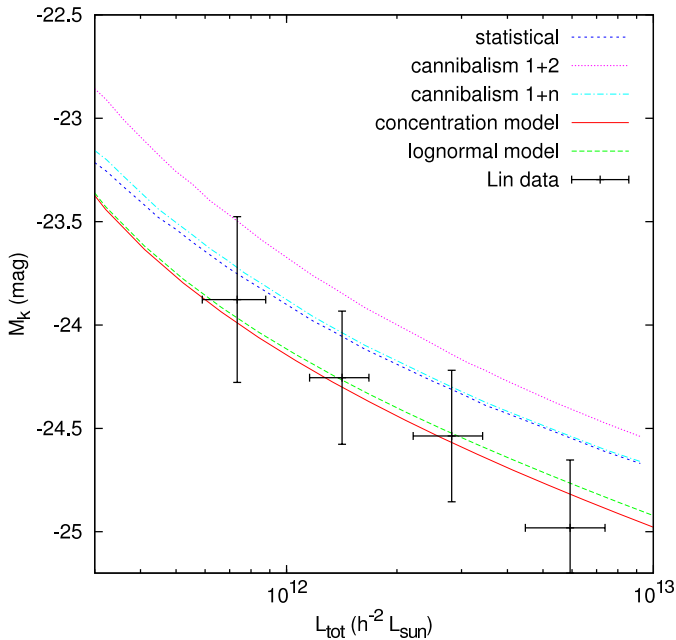}
\caption{Same as figure \ref{m1avgfig}, but for the second brightest
galaxy in the cluster.}
\label{m2avgfig}
\end{center}
\end{figure}

We start by showing the results for average magnitude of first and
second brightest galaxies as a function of total cluster luminosity,
for each of the models, in figures \ref{m1avgfig} and
\ref{m2avgfig}. The first noticeable thing is that the curves for
both the concentration and lognormal models are very similar, which
is not too surprising given the similar mass-luminosity relations
obtained for each of them (see table 2), even though they were built
in independent ways. This is in fact a success for the concentration
model, since the lognormal model is by construction made to provide 
the best fit to the observed BCG magnitude, while the concentration
model is not.

Comparing the different sets of curves, it is easy to see the differences
between the models. The statistical distribution has the lowest
average BCG luminosity, which is considerably higher for the other 
models. This comes from the fact that all other models take the BCG
distribution to be special, like an additional distribution added on
to the base Schechter distribution of the satellite galaxies. Looking
at the average magnitude of the second brightest galaxy, the most
obvious factor is that the one from the more extreme cannibalism model is much
lower than the others; once again, this is unsurprising since this
is essentially the third brightest galaxy in the statistical
model. Likewise, the 1+n cannibalism model gives essentially the same 
result as the statistical one, since the second brightest galaxy is
the same in both in most cases. Both of our models give considerably
brighter values for the second brightest galaxy. This probably indicates
that the luminosity function we used to generate the statistical and 
cannibalism models does not have enough bright galaxies (i.e., $L_*$
is too low), which is not very surprising since we are using the parameters
of the global one. On the other hand, if this were to be the case
then it is quite probable that the cannibalism models would then 
give BCGs which are too bright.

\begin{figure}
\begin{center}
\includegraphics[width=84mm,angle=270]{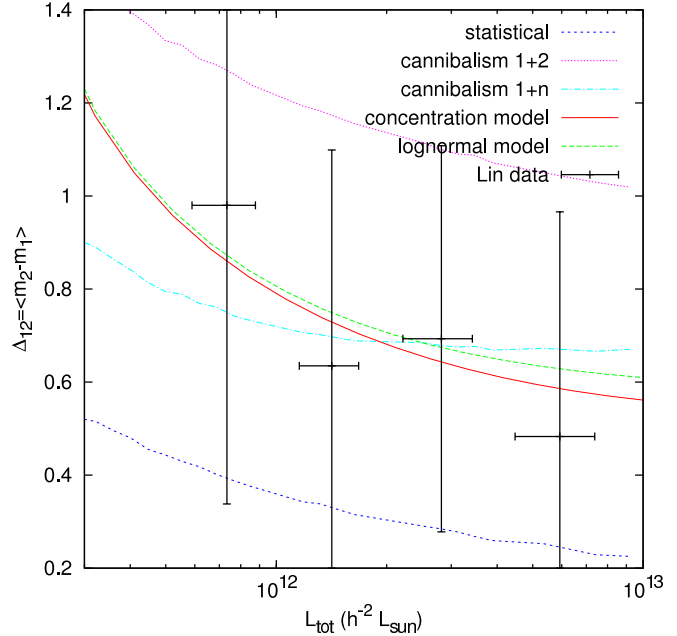}
\caption{Values for the average magnitude difference between first and
second brightest galaxies, $\Delta_{12}$. The data points come are again binned 
values from the catalogue supplied by Yen-Ting Lin. The large errorbars reflect the
fact that the scatter in the values of $\Delta_{12}$ in these clusters is very large.
For comparison, another observational value for 
$\Delta_{12}$, but in the $B$-band, is of
$\approx 0.55$ magnitudes \citep{sgh}. The values we obtain are also
qualitatively consistent with the results of \citet{milos,ls}.}  
\label{delta12fig} 
\end{center}
\end{figure}

In any case, this problem should be less of an issue when looking 
at the magnitude gap $\Delta_{12}$, shown
in figure \ref{delta12fig}. The values for the cannibalism 1+2
model are obviously much too high to match the observed ones: it is
clear that merging the two brightest galaxies leaves too big a gap to
the next brightest. The values for the statistical model are too low,
in this case probably indicating that it is the BCG which is too faint.
The values for the other, 1+n, cannibalism model look rather better, 
while both of our models show good agreement with the observations.
But note the size of the errorbars in the observational data: the 
scatter in the observed values is quite large (see also \citealt{ls,vdb}).
Since the scatter in the BCG magnitude is small, most of the scatter here
in $\Delta_{12}$ is likely coming from the second brightest galaxy.
The shape seen in the curves of
figure \ref{delta12fig} is unsurprising: the fact that $\Delta_{12}$
is increasing as the total luminosity goes down is a natural
consequence of the decreasing number of satellites at this lower
end. In fact, these correspond to haloes of only a few times $10^{13}
h^{-1} {\rm M_\odot}$, and therefore many of these systems will not
actually even be clusters (cf figure \ref{clustprob}).

\begin{figure}
\begin{center}
\includegraphics[width=84mm,angle=270]{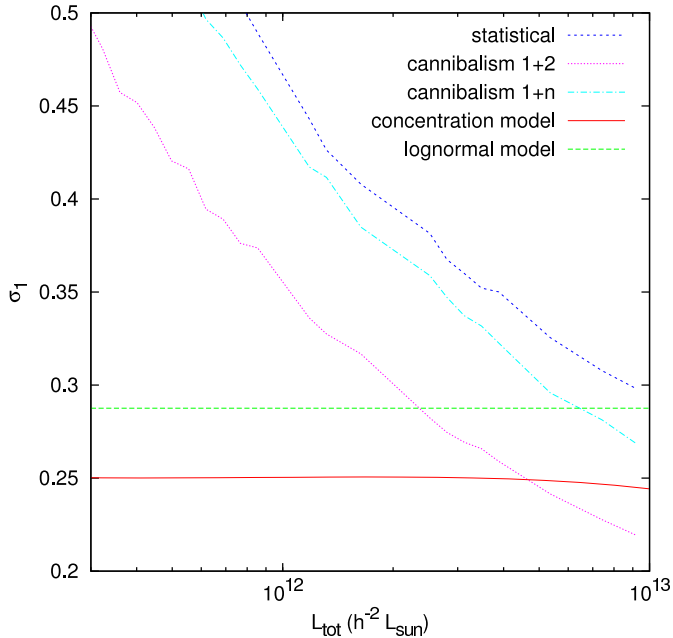}
\caption{Dispersion in the magnitude of the first brightest galaxy. 
The data set used to plot the data points in the previous
figures gives a value of between 0.3 and 0.4 magnitudes for the statistical
error in each cluster luminosity bin. Another observationally 
measured dispersion in first brightest
magnitude, in the $B$-band, is $\sigma_1\simeq0.24$ \citep{pl}. 
Using a CLF approach similar to ours, \citet{coorayc}
obtains a value of $\sigma_1=0.17$ in the r-band.}
\label{sigfig}
\end{center}
\end{figure}

Figure \ref{sigfig} shows the calculated dispersion in the magnitude
of the first brightest galaxy. Binning the Lin observational data 
results in values of $\sigma_1$ of 0.3 to 0.4 magnitudes. 
A direct comparison of this value
with the model results shows that the values we obtain for our model
are slightly too low, while the values for the statistical and
cannibalism models look reasonable. There is, however, an additional
point to bear in mind: these observational values are obtained by binning
over a certain cluster luminosity range. 
Part of this observed scatter then simply comes from the fact that 
the average magnitude is changing as a function of this. When taking this
into account, the values we get from our model are in fact in good
agreement with the observed ones. Also, using Bayes' theorem, 
it is possible to derive values for the dispersion in the mass 
of the hosting halo for a given galaxy luminosity. Doing so we obtain
values which are in qualitative agreement with the ones found by 
\citet{vdb}, although a direct comparison is complicated by the
fact that they use $b_J$-band instead of $K$-band for the galaxy
luminosity.

\section{Conclusion}

In the present paper we focus on the implications of the mass
luminosity relation first introduced in paper I on the properties of
clusters, and more in particular on whether BCGs are statistical or
special in nature. We expand the model of paper I by introducing
scatter into the relation, building a CLF based on two different
models: a simple lognormal shape or a model based on the distribution
of concentrations in haloes of a given mass. We also introduce a 
simple model to evaluate the mass fraction in subhaloes which have
been completely disrupted since they were accreted.

We have argued that this model naturally gives a separation between
central and satellite galaxies in a cluster, with the former having a
distinct distribution based on the halo mass function. We have shown
that this leads to a characteristic bump in the cluster galaxy
luminosity function, qualitatively similar to that seen in some
observational work and some semi-analytical HOD based models. This is
caused by the fact that, at any given high mass, haloes are
considerably more abundant than subhaloes, coupled with the fact that,
in any single system, central galaxies will be considerably brighter
than the satellites (since the subhaloes are much less massive than
the parent halo). Together with this, the faint end is completely
determined by the satellite galaxies in the subhaloes, since we put in
a minimum mass threshold for the haloes we consider host clusters.

Finally, to look at the question of the nature of BCGs, we
 study some statistical indicators that may provide a clue
to this problem, namely the ratio $r$ between the magnitude difference
between first and second brightest galaxies in a cluster,
$\Delta_{12}$, and the dispersion of the former, $\sigma_1$. We also
introduce two simple models to account for two different possibilities
that are usually considered: the statistical hypothesis, that is, that
all galaxies in a cluster, including the BCG, are drawn from the same
distribution; and galactic cannibalism, where the BCG grows by merging
with other galaxies in the cluster. As is already known, any model of
the former gives a value for $r$ which is too small compared to what
is observed. At the same time, we show that the simplest case of the
latter, that of merging the two brightest galaxies from a common
distribution, gives a value of $\Delta_{12}$ which is far too large.

From the results we obtain, it is possible to draw some answers to the
issue discussed in the introduction about the nature of BCGs. The
statistical hypothesis, which assumes that BCGs are drawn from the
same universal distribution, can be ruled out. It gives values of
$\sigma_1$ which are too big and $\Delta_{12}$ which are too small. 

On the other hand, the simplistic model we analyse for galactic
cannibalism looks to be far too extreme. Mainly, it gives values of
$\Delta_{12}$ which are far too large compared to what is
observed. This model mimics the scenario of two similar sized clusters
merging, with a final distribution of galaxies which can be well fit
by a single distribution, but where the BCG is then built up by
merging the two BCGs of the original clusters. From our results, we
expect that if such a scenario does occur, a merging of the brightest
galaxies is excluded, as it leaves a too bright BCG and too large a
gap to the second brightest galaxy (but see the discussion in
\citealt{lm}). A more general cannibalistic scenario, where the BCG is
built up by merging with one other galaxy, is however not excluded,
nor is the possibility of minor mergers, where one of the merging
systems is much smaller than the other such that its brightest galaxy
will not be the second brightest galaxy in the resulting cluster. It 
is worth noting that this fits in well with the results of 
recent semi-analytic simulations done by 
\citet{lucia}, who find that BCG growth occurs fast enough that major
mergers are relatively rare, and that most of the later growth is 
through minor mergers.

Our model gives results which can be regarded as halfway between the
other two: the BCG is in fact the product of a special distribution
and is considerably brighter than what would result from a single
distribution, but the second brightest galaxy is not as faint as in
the cannibalistic scenario. The question remains,
however, of what is the BCG formation scenario expected in this
case. A satisfactory answer, in the framework of the way the mass
luminosity relation is built, would require going back to the
simulations and following the behaviour of the halo and its
subhaloes. 

There is an important assumption which has been implicit throughout
this paper: that both the central and satellite galaxies follow the
same mass luminosity relation. This may indeed appear to be
contradictory with the possibility that the central galaxies are
special, since it may seem to imply that they are formed through the
same processes. On the other hand, the model is based on structure
build up through the merger of dark matter haloes, and we assume that
the satellite galaxies were formed in their own independent haloes,
prior to being accreted into their present parent halo. This original
halo would be the one that determines their properties, since the
galaxy stops accreting gas or undergoing mergers of its own once its
halo merges into the parent system and it becomes a satellite (and
hence the need for some mass loss prescription). Therefore, it is to
be expected that at least in some cases, they would have been a BCG in
their own system themselves, and therefore it may not be unreasonable
to assign them the same mass luminosity relation. Still, this leaves
out the very important factor that the formation epoch may well be
different, together with subsquent growth of the BCG in the parent
system. At the same time, halo numbers completely dominate at high
mass (and therefore it is to be expected the same is true of central
galaxy numbers at high luminosity), and consequently the total average
mass luminosity relation should be pretty much the same as
calculated. 

We should stress the fact that our concentration models gives quite
good results, since unlike the lognormal model, it does not involve
any fitting of parameters to BCG results. In fact, it is quite interesting
that it gives values for the dispersion in BCG magnitude so close to 
the ones from the lognormal model, when the latter was set to the
value that results in the best fit to the observed BCG magnitudes.
In light of the recent discussion about the effects of additional
parameters, such as environmental density, halo formation time or
concentration (e.g., \citealt{wechsler,berlind}), 
on the clustering of halos and subsequently on the
HOD, this seems to indicate that taking the halo mass as the primary
determinant of the hosted galaxy luminosity, and then taking 
concentration as a secondary variable which determines the scatter 
around the average value is a good way to proceed.

\section*{Acknowledgements}
We would like to thank Yen-Ting Lin for making his data on 
cluster galaxy luminosities available to us.
AV acknowledges financial support from Funda\c
c\~ao para a Ci\^encia e Tecnologia (Portugal), under grant
SFRH/BD/2989/2000.


\begin{thebibliography}{99}
\bibitem[\protect\citeauthoryear{Benson et al.}{2003a}]{bfb}
Benson A.~J., Frenk C.~S., Baugh C.~M., Cole S., Lacey C.~G.,
2003, MNRAS, 343, 679
\bibitem[\protect\citeauthoryear{Bhavsar}{1989}]{bhavsar}
Bhavsar S.P., 1989, ApJ, 338, 718
\bibitem[\protect\citeauthoryear{Bhavsar \& Barrow}{1985}]{bb84}
Bhavsar S.P., Barrow J.D., 1985, MNRAS, 213, 857
\bibitem[\protect\citeauthoryear{Berlind \& Weinberg}{2002}]{bg}
Berlind A.~A., Weinberg D.~H., 2002, ApJ, 575,587
\bibitem[\protect\citeauthoryear{Berlind et al.}{2006}]{berlind}
Berlind A.A., Kazin E., Blanton M.R., Pueblas S., Scoccimarro R., Hogg D.W.,
astro-ph/0610524, submitted to ApJ
\bibitem[\protect\citeauthoryear{Bernardi et al.}{2006}]{bernardi}
Bernardi M., Hyde J.B., Sheth R.K., Miller C.J., Nichol R.C., 2006, ApJ, in press, astro-ph/0607117
\bibitem[\protect\citeauthoryear{Bernstein \& Bhavsar}{2000}]{bb}
Bernstein J.P., Bhavsar S.P., 2000, MNRAS
\bibitem[\protect\citeauthoryear{Binggeli, Sandage \& Tammann}{1988}]{cdreview} 
Binggeli B., Sandage A., Tammann G.A., 1988, ARA\&A, 26, 509
\bibitem[\protect\citeauthoryear{Bullock et al.}{2001}]{bullock}
Bullock, J.~S., Kolatt, T.~S., Sigad, Y., Somerville, R.~S., Kravtsov, A.~V., Klypin, A.~A., 
Primack, J.~R., Dekel, A., 2001, MNRAS, 321, 559 
\bibitem[\protect\citeauthoryear{Capaccioli}{1989}]{capaccioli}
Capaccioli M., 1989, in The World of Galaxies, ed. H.G. Corwin\&
L. Bottinelli, Springer, Berlin, 208
\bibitem[\protect\citeauthoryear{Colless}{1989}]{colless} 
Colless M., 1989, MNRAS, 237, 799 
\bibitem[\protect\citeauthoryear{Conroy, Wechsler \& Kravtsov}{2006}]{charlie}
Conroy C., Wechsler R.H., Kravtsov A.V., 2006, ApJ, 647, 201
\bibitem[\protect\citeauthoryear{Cooray}{2006}]{coorayc}
Cooray A., 2006, MNRAS, 365, 842
\bibitem[\protect\citeauthoryear{Cooray \& Milosavljevi\'c}{2005a}]{cooray}
Cooray A., Milosavljevi\'c M., 2005a, ApJ, 627, L85
\bibitem[\protect\citeauthoryear{Cooray \& Milosavljevi\'c}{2005b}]{coorayb}
Cooray A., Milosavljevi\'c M., 2005b, ApJ, 627, L89
\bibitem[\protect\citeauthoryear{de Lucia \& Blaizot}{2006}]{lucia}
De Lucia G., Blaizot J., 2006, MNRAS, accepted, astro-ph/0606519
\bibitem[\protect\citeauthoryear{de Propris et al.}{2003}]{2dfcluster}
De Propris R.~et al., 2003, MNRAS, 342, 725 
\bibitem[\protect\citeauthoryear{Dressler}{1978}]{dressler} 
Dressler A., 1978, ApJ, 222, 23 
\bibitem[\protect\citeauthoryear{Eke et al.}{2004}]{2pigg}
Eke V.R. et al., 2004, MNRAS, 355, 769
\bibitem[\protect\citeauthoryear{Gao et al.}{2004}]{gao}
Gao L., White S.D.M., Jenkins A., Stoehr F., Springel V., MNRAS, 355, 819
\bibitem[\protect\citeauthoryear{Graham et al.}{1996}]{graham}
Graham A., Lauer T.R., Colless M., Postman M., 1996, ApJ, 465, 534
\bibitem[\protect\citeauthoryear{Hausman \& Ostriker}{1978}]{ho}
Hausman M.A., Ostriker J.P., 1978, ApJ, 224, 320
\bibitem[\protect\citeauthoryear{Hoessel, Gunn \& Thuan}{1980}]{hgt}
Hoessel J.G., Gunn J.E., Thuan T.X., 1980, ApJ, 241, 486
\bibitem[\protect\citeauthoryear{Hoessel \& Schneider}{1985}]{hs}
Hoessel J.G., Schneider D.P., 1985, AJ, 90, 1648
\bibitem[\protect\citeauthoryear{Kochanek et al.}{2001}]{2mass}
Kochanek C.~S., et al., 2001, ApJ, 560, 566 
\bibitem[\protect\citeauthoryear{Kravtsov et al.}{2004}]{kravtsov}
Kravtsov A.~V., Berlind A.~A., Wechsler R.~H., Klypin A.~A.,
Gottl{\"o}ber A., Allgood B., Primack J.~R., 2004, ApJ, 609, 35
\bibitem[\protect\citeauthoryear{Lin \& Mohr}{2004}]{lm}
Lin Y., Mohr J.J., 2004, ApJ, 617, 879
\bibitem[\protect\citeauthoryear{Lin, Mohr \& Stanford}{2004}]{lms}
Lin Y., Mohr J.J., Stanford S.A., 2004, ApJ, 610, 745
\bibitem[\protect\citeauthoryear{von der Linden et al.}{2006}]{linden}
von der Linden A., Best P.N., Kauffmann G., White S.D.M., 2006, astro-ph/0611196,
submitted to MNRAS
\bibitem[\protect\citeauthoryear{Loh \& Strauss}{2006}]{ls}
Loh Y., Strauss M.A., 2006, MNRAS, 366, 373
\bibitem[\protect\citeauthoryear{Macci\`o et al.}{2006}]{maccio}
Macci\`o A.V., Dutton A.A., van den Bosch F.C., Moore B., Potter D., Stadel J., 2006,
astro-ph/0608157, submitted to MNRAS
\bibitem[\protect\citeauthoryear{Milosavljevi\'c et al.}{2006}]{milos}
Milosavljevi\'c M., Miller C.J., Furlanetto S.R., Cooray A., 2006, ApJ, 637, L9
\bibitem[\protect\citeauthoryear{Navarro, Frenk, \& White}{1997}]{nfw}
Navarro J.~F., Frenk C.~S., White S.~D.~M., 1997, ApJ, 490, 493
\bibitem[\protect\citeauthoryear{Norberg et al.}{2002}]{2df}
Norberg P.~et al., 2002, MNRAS, 336, 907
\bibitem[\protect\citeauthoryear{Ostriker \& Hausman}{1977}]{oh}
Ostriker J.P., Hausman M. A., 1977, ApJ, 217, L125
\bibitem[\protect\citeauthoryear{Ostriker \& Tremaine}{1975}]{ot}
Ostriker J.P., Tremaine S.D., 1975, ApJ, 202, 113
\bibitem[\protect\citeauthoryear{Peebles}{1968}]{peebles}
Peebles, P.J.E., 1968, ApJ, 153, 13
\bibitem[\protect\citeauthoryear{Postman \& Lauer}{1995}]{pl}
Postman M., Lauer T.R., 1995, ApJ, 440, 28
\bibitem[\protect\citeauthoryear{Sandage}{1972}]{sandage}
Sandage, A., 1972, ApJ, 178, 1
\bibitem[\protect\citeauthoryear{Schneider, Gunn, \& Hoessel}{1983}]{sgh} 
Schneider D.~P., Gunn J.~E., Hoessel J.~G., 1983, ApJ, 268, 476 
\bibitem[\protect\citeauthoryear{Shaw et al.}{2006}]{laurie}
Shaw L., Weller J., Ostriker J.P., Bode P., 2006, ApJ, 646, 815
\bibitem[\protect\citeauthoryear{Sheth \& Tormen}{1999}]{stmf}
Sheth R.~K., Tormen G., MNRAS, 1999, 308, 119
\bibitem[\protect\citeauthoryear{Spergel et al.}{2006}]{wmap}
Spergel D.~N.~et al., 2006, astro-ph/0603449, submitted to ApJ
\bibitem[\protect\citeauthoryear{Tasitsiomi et al.}{2004}]{iro}
Tasitsiomi A., Kravtsov A.~V., Wechsler R.~H., Primack J.~R., 2004, ApJ. 614, 533
\bibitem[\protect\citeauthoryear{Tremaine \& Richstone}{1977}]{tr}
Tremaine S.D., Richstone D.O., 1977, ApJ, 212, 311
\bibitem[\protect\citeauthoryear{Vale \& Ostriker}{2004}]{paper1} 
Vale A., Ostriker J.~P., 2004, MNRAS, 353, 189 
\bibitem[\protect\citeauthoryear{Vale \& Ostriker}{2006}]{paper2} 
Vale A., Ostriker J.~P., 2006, MNRAS, 371, 1173 (Paper I)
\bibitem[\protect\citeauthoryear{van den Bosch, Tormen \& Giocoli}{2005}]{vdbshmf}
van den Bosch F.~C., Tormen G., Giocoli C., 2005, MNRAS, 359, 1029
\bibitem[\protect\citeauthoryear{van den Bosch et al.}{2006}]{vdb}
van den Bosch F.C. et al., 2006, astro-ph/0610686, submitted to MNRAS
\bibitem[\protect\citeauthoryear{Wang et al.}{2006}]{wang}
Wang L., Li C., Kauffmann G., De Lucia G., 2006, MNRAS, 371, 537
\bibitem[\protect\citeauthoryear{Wechsler et al.}{2006}]{wechsler}
Wechsler R.H., Zentner A.R., Bullock J.S., Kravtsov A.V., 2006, ApJ, in press, astro-ph/0512416
\bibitem[\protect\citeauthoryear{Weller et al.}{2005}]{jochen} 
Weller J., Ostriker J.~P., Bode P., Shaw L., 2005, MNRAS, 364, 823
\bibitem[\protect\citeauthoryear{Yagi et al.}{2002}]{yagi} 
Yagi M., Kashikawa N., Sekiguchi M., Doi M., Yasuda N., Shimasaku K., 
Okamura S., 2002, AJ, 123, 87 
\bibitem[\protect\citeauthoryear{Yang et al.}{2003}]{yang}
Yang X.H., Mo H.J., van den Bosch F.C., 2003, MNRAS, 339, 1057
\bibitem[\protect\citeauthoryear{Yang et al.}{2005}]{yanghod}
Yang X.H., Mo H.J., Jing Y.P., van den Bosch F.C., 2005, MNRAS, 358, 217
\bibitem[\protect\citeauthoryear{Zehavi et al.}{2005}]{zehavi}
Zehavi I. et al., 2005, ApJ, 630, 1
\bibitem[\protect\citeauthoryear{Zentner et al.}{2005}]{zentner}
Zentner A.R., Berlind A.A., Bullock J.S., Kravtsov A.V., Wechsler R.H., 2005, 
ApJ, 624, 505
\bibitem[\protect\citeauthoryear{Zheng et al.}{2005}]{zz}
Zheng Z. et al., 2005, ApJ, 633, 791
\bibitem[\protect\citeauthoryear{Zhu et al.}{2006}]{zhu}
Zhu G., Zheng Z., Lin, W.P., Jing Y.P., Kang X., Gao L., 2006, astro-ph/0601120, 
submitted to ApJ
\end{thebibliography}
\end{document}